\documentclass[iop]{piner3}
\usepackage{myrotating}
\begin{document}

\shorttitle{Higher-Resolution Observations of TeV Blazars}

\title{The Jets of TeV Blazars at Higher Resolution: 
43 GHz and Polarimetric VLBA Observations from 2005-2009}

\author{B.~Glenn~Piner\altaffilmark{1}, Niraj~Pant\altaffilmark{1}, \& Philip~G.~Edwards\altaffilmark{2}}

\altaffiltext{1}{Department of Physics and Astronomy, Whittier College,
13406 E. Philadelphia Street, Whittier, CA 90608; gpiner@whittier.edu}

\altaffiltext{2}{CSIRO Astronomy and Space Science,
PO Box 76, Epping NSW 1710, Australia; Philip.Edwards@csiro.au}

\begin{abstract}
We present 23 new VLBA images of the six established TeV blazars Markarian 421, Markarian 501, 
H~1426+428, 1ES~1959+650, PKS~2155$-$304, and 1ES~2344+514, obtained from 2005 to 2009.
Most images were obtained at 43 GHz (7~mm), and they reveal the parsec-scale structures of three of
these sources (1ES~1959+650, PKS~2155$-$304, and 1ES~2344+514)
at factors of two to three higher resolution than has previously been attained.
These images reveal new morphological details, including a high degree of jet bending in the
inner milliarcsecond in PKS~2155$-$304.
This establishes strong apparent jet bending on VLBI scales as a common property of TeV blazars,
implying viewing angles close to the line-of-sight.
Most of the remaining images map the linear polarization structures at a lower frequency
of 22 GHz (1~cm). We discuss the transverse structures of the jets as revealed by the high-frequency and
polarimetric imaging. The transverse structures include significant limb-brightening in Mrk~421,
and `spine-sheath' structures in the electric vector position angle (EVPA) and fractional polarization distributions in
Mrk~421, Mrk~501, and 1ES~1959+650.
We use new measured component positions to update measured apparent jet speeds, in many cases
significantly reducing the statistical error over previously published results.
With the increased resolution at 43~GHz, we detect new components within 0.1-0.2 mas of the core
in most of these sources. No motion is apparent in these new components over the time span of our observations, and
we place upper limits on the apparent speeds of the components near the core of 
$<2c$. From those limits, we conclude that $\Gamma_{2}<(\Gamma_{1})^{1/2}$ at $\sim10^{5}$ Schwarzschild radii, where
$\Gamma_{1}$ and $\Gamma_{2}$ are the bulk Lorentz factors in the TeV-emitting and 43~GHz-emitting
regions, respectively, assuming that their velocity vectors are aligned.
\end{abstract}


\section{Introduction}
\label{intro}
The latest generation of TeV gamma-ray telescopes have 
enabled detections of increasing numbers of TeV sources over the past several years, as well
as improved measurements of spectra and variability in previously established sources.
Most of the extragalactic sources detected at these TeV energies (currently 23 of 36 sources
\footnote{tevcat.uchicago.edu}) are high-frequency-peaked
BL Lac objects, or HBLs, a subclass of blazar whose synchrotron and inverse-Compton
spectra peak at high frequencies. The blazar phenomenon as a whole is now well understood to
result from a relativistically moving jet of plasma directed nearly at the observer, but
the TeV HBLs in particular have presented a number of interesting problems.

The strong variability found in at least two of these blazars:
PKS~2155$-$304 (Aharonian et al. 2007) and Mrk~501 (Albert et al. 2007), on time-scales as
short as a few minutes (considerably shorter than the shortest time-scales expected),
provoked considerable interest.
Modeling of this variability has implied bulk Lorentz factors and Doppler factors exceeding 50
(Begelman et al. 2008; Finke et al. 2008), 
even extending up to $\sim100$ (Kusunose \& Takahara 2008).
However, other authors employing specific time-dependent or inhomogeneous jet models
reproduce the spectra and short variability with lower bulk Lorentz factors of about 15 to 30
(Boutelier et al. 2008; Katarzy\'{n}ski et al. 2008). Some authors have proposed decoupling
the flare emission from the bulk motion entirely (Ghisellini et al. 2009), by having
the flare produced by aligned beams of particles moving at the particle Lorentz factor.
Even though the most extreme values have usually been invoked to model the variability,
moderately high Doppler factors are required to reproduce the Spectral Energy Distributions (SEDs)
of the TeV blazars in the context of one-zone synchrotron self-Compton models. For example,
Tavecchio et al. (2010) find Doppler factors in the range of 20 to 40 when modeling the
SEDs of the TeV blazars, including the MeV--GeV data for those TeV blazars detected by {\em Fermi}.

A major complementary observing technique to use in understanding the 
physics of the TeV blazars is very long baseline interferometry (VLBI).
VLBI reveals the structure of the parsec-scale jets through direct imaging, 
rather than from timing or multiwavelength spectral information.
We have been pursuing this strategy with the National Radio Astronomy Observatory's
Very Long Baseline Array (VLBA) since the discovery of the TeV blazars.
As part of this program, we have previously studied the parsec-scale structure of Mrk~421 (Piner et al. 1999; Piner \&
Edwards 2005), Mrk~501 (Edwards \& Piner 2002; Piner et al. 2009, hereafter P09),
and the four sources H~1426+428, 1ES~1959+650, PKS~2155$-$304, and 1ES~2344+514
(Piner \& Edwards 2004; Piner et al. 2008, hereafter P08).

Along with other workers studying HBLs in the radio (e.g., Giroletti et al. 2004a, 2006; Wu et al. 2007),
we have consistently found that the bulk Lorentz factors in the parsec-scale radio jets of TeV HBLs are
fairly modest, $\Gamma\sim3$ to 4 (see the discussion of this point in P08, and
the discussion of this point for a larger set of HBLs, including the TeV sources, in Giroletti et al. 2004a, 2006).
Similarly modest bulk Lorentz factors are required by the unification of HBLs and FR~I radio galaxies 
(Chiaberge et al. 2000). The difference between the high Lorentz factors derived from the TeV emission
(both from SED modeling and from the variability),
and the modest Lorentz factors found from the compact radio emission and from unification has been dubbed
the `bulk Lorentz factor crisis' (Henri \& Saug\'{e} 2006) of TeV blazars. 

The bulk Lorentz factor crisis suggests that the TeV and radio emission are produced
in regions of different Lorentz factor, although it is not clear how these two regions may be
arranged, and what underlying physics might be responsible. 
The slower region could be located downstream of the faster region, implying a 
decelerating jet (e.g., Georganopoulos \& Kazanas 2003; 
Levinson 2007; Stern \& Poutanen 2008); note that a decelerating
jet also makes sense from energy-budget considerations for this 
particular class of sources (Celotti \& Ghisellini 2008). The slower region could surround a faster
central region in a `spine-sheath' structure (Ghisellini et al. 2005).
Such a two component outflow is expected from theoretical models of jet launching 
(Meliani et al. 2010 and references therein). 
Under certain conditions, the transverse profile of the Lorentz factor may even
be nonmonotonic (Aloy \& Mimica 2008). Some of the above models
to resolve the bulk Lorentz factor crisis (e.g., Stern \& Poutanen 2008; Ghisellini et al. 2005)
also naturally include {\em both} longitudinal and transverse gradients in the bulk Lorentz factor.

Alternatively, the fast region could be a smaller sub-structure propagating inside a larger, slower jet.
At least three models propose specific variations of this idea: the `needle/jet' model by
Ghisellini \& Tavecchio (2008), the `jet in a jet' model by Giannios et al. (2009), and
the model by Lyutikov \& Lister (2010) where the leading edge of a non-stationary ejection
in a highly magnetized flow is accelerated to much higher Lorentz factors than the slower-moving bulk flow.
There is also an alternate class of solution (e.g., Gopal-Krishna et al. 2007 and references therein)
where the slow apparent speeds measured
by VLBI result from a broad conical jet viewed almost along the line-of-sight, but this 
does not resolve the several other indicators of modest beaming in the radio, apart from the
apparent jet speeds (see the discussion of this model in P08).
Any longitudinal or transverse velocity structures that exist will in turn affect the computed SEDs
(e.g., Yang et al. 2009), so any such structures are important to take into account.

Because either transverse or longitudinal jet structures in the TeV HBLs are a common feature of
proposed solutions to the bulk Lorentz factor crisis,
the goals of the VLBA observations presented in this paper were to both make
detailed observations of transverse structures perpendicular to the jet axes, and to observe the
structures closer to the VLBI core than had been possible in our previous observations.
Because of the need for higher resolution, the majority of the images in this paper are at 43~GHz,
as compared to a typical frequency of 15~GHz used in most of our earlier studies of these sources.
At 43~GHz, the resolution of the VLBA is about 0.1~pc at the distance of the closest TeV blazars,
or about 1,000 Schwarzschild radii for a $10^{9}$ solar mass black hole.
Many of the newer and fainter TeV HBLs are below the VLBA's detection limit at 43~GHz; for this reason this paper
concentrates on the six earliest established (and typically radio brightest) TeV HBLs. 
Even among these sources, a high data rate (for that time)
of 512 Mbps was required for the VLBA observations, and some sources
were near the VLBA's detection limit.

Some of these data have been incorporated in various multiwavelength campaigns; their use in the context of the various
multiwavelength campaigns will be described in those papers. 
Any discussions of radio core variability compared to TeV flux variability will also be presented there.
A complementary paper on lower-frequency multi-epoch VLBA observations of the more recently detected
(and radio fainter) TeV blazars is also in preparation.

The organization of this paper is as follows: in $\S$~\ref{obs} we describe the details of 
our VLBA observations, in $\S$~\ref{results} we present the imaging and model fitting results for specific sources,
in $\S$~\ref{discussion} we discuss the physical results obtained for the entire sample,
and in $\S$~\ref{conclusion} we present our major conclusions.
Throughout the paper, we assume cosmological parameters of 
$H_{0}=71$ km s$^{-1}$ Mpc$^{-1}$, $\Omega_{m}=0.27$, and $\Omega_{\Lambda}=0.73$.
When results from other papers are quoted,
they have been converted to this cosmology.

\section{Observations}
\label{obs}
The VLBA observations considered in this paper are drawn from three VLBA experiments:
BP112 during 2005, BE044 during 2006 and 2007, and BP143 during 2008 and 2009.

VLBA experiment BP112 recorded high-frequency 43 and 86~GHz observations of Mrk~421 and Mrk~501
at three epochs during 2005, at a high data rate of 512~Mbps. 
The BP112 results on Mrk~501 have already been reported elsewhere (P09).
Both of these sources were near the VLBA's detection limit at 86~GHz; Mrk~501 was detected
at its brightest epoch, but Mrk~421 was not detected at 86~GHz.
However, images were produced of Mrk~421 at 43~GHz at each epoch.
Details of the three 43 GHz observations of Mrk~421 from VLBA experiment BP112 are given in Table~\ref{imtab},
further details on the reduction of experiment BP112 are given in P09.

\begin{sidewaystable*}[!t]
\caption{Observation Log and Parameters of the Images}
\vspace{-0.15in}
\begin{center}
\label{imtab}
{\scriptsize \begin{tabular}{l l l l c c c l c c c} \colrule \colrule \\ [-9pt]
& & & & & Time On & & & Peak Flux \\
& & \multicolumn{1}{c}{Observing} & \multicolumn{1}{c}{VLBA} & Frequency & Source &
& & Density & $I_{rms}^{d}$ & $P_{rms}^{d}$ \\
\multicolumn{1}{c}{Source} & \multicolumn{1}{c}{Epoch} & \multicolumn{1}{c}{Code} &
\multicolumn{1}{c}{Antennas$^{a}$} & (GHz) & (hr) & Pol$^{b}$ &
\multicolumn{1}{c}{Beam$^{c}$} & (mJy bm$^{-1}$) & (mJy bm$^{-1}$) & (mJy bm$^{-1}$) \\ \colrule \\ [-5pt]
Markarian 421  & 2005 May 21 & BP112A & No HN,SC & 43.2 & 0.5 & No  & 0.51,0.19,$-$18.6 & 164 & 0.42 & ...  \\
               & 2005 Jun 21 & BP112B & No HN,SC & 43.2 & 0.5 & No  & 0.48,0.29,$-$18.8 & 127 & 0.72 & ...  \\
               & 2005 Sep 2  & BP112C & No HN,SC & 43.2 & 0.5 & No  & 0.51,0.17,$-$15.6 & 174 & 0.41 & ...  \\
               & 2006 Oct 15 & BE044A & No BR    & 22.2 & 3.5 & Yes & 0.89,0.33,$-$0.3  & 234 & 0.28 & 0.22 \\
               & 2008 Dec 4  & BP143A & No SC    & 43.2 & 1.0 & No  & 0.49,0.20,$-$37.2 & 193 & 0.18 & ...  \\
               & 2009 Jan 10 & BP143B & No HN,SC & 43.2 & 1.0 & No  & 0.51,0.19,$-$31.2 & 198 & 0.21 & ...  \\
               & 2009 Feb 12 & BP143C & No SC    & 43.2 & 1.0 & No  & 0.49,0.24,$-$31.0 & 149 & 0.21 & ...  \\
H~1426+428     & 2006 Oct 22 & BE044E & All      & 8.4  & 4.5 & No  & 2.00,0.92,$-$5.8  & 18  & 0.05 & ...  \\
Markarian 501  & 2006 Oct 16 & BE044B & No BR    & 22.2 & 3.5 & Yes & 0.96,0.34,$-$12.6 & 503 & 0.60 & 0.25 \\
               & 2008 Dec 4  & BP143A & All      & 43.2 & 1.0 & No  & 0.43,0.20,20.0    & 256 & 0.38 & ...  \\
               & 2009 Jan 10 & BP143B & No HN,NL & 43.2 & 1.0 & No  & 0.47,0.20,26.6    & 262 & 0.37 & ...  \\
               & 2009 Feb 12 & BP143C & All      & 43.2 & 1.0 & No  & 0.44,0.22,26.4    & 229 & 0.31 & ...  \\
1ES~1959+650   & 2007 Jan 27 & BE044C & No KP    & 22.2 & 3.5 & Yes & 0.70,0.33,2.4     & 95  & 0.18 & 0.14 \\
               & 2008 Dec 4  & BP143A & All      & 43.2 & 2.0 & No  & 0.47,0.22,49.4    & 62  & 0.17 & ...  \\
               & 2009 Jan 10 & BP143B & No HN    & 43.2 & 2.0 & No  & 0.54,0.24,59.4    & 57  & 0.16 & ...  \\
               & 2009 Feb 12 & BP143C & All      & 43.2 & 2.0 & No  & 0.51,0.27,51.4    & 60  & 0.18 & ...  \\
PKS~2155$-$304 & 2008 Dec 4  & BP143A & All      & 43.2 & 2.0 & No  & 0.60,0.19,0.3     & 120 & 0.26 & ...  \\
               & 2009 Jan 10 & BP143B & No HN    & 43.2 & 2.0 & No  & 0.65,0.20,2.4     & 86  & 0.33 & ...  \\
               & 2009 Feb 12 & BP143C & No BR    & 43.2 & 2.0 & No  & 0.67,0.19,1.6     & 112 & 0.37 & ...  \\
1ES~2344+514   & 2006 Oct 17 & BE044D & No BR    & 22.2 & 3.5 & Yes & 0.93,0.39,$-$20.7 & 107 & 0.23 & 0.22 \\
               & 2008 Dec 4  & BP143A & All      & 43.2 & 1.5 & No  & 0.54,0.26,61.3    & 36  & 0.24 & ...  \\
               & 2009 Jan 10 & BP143B & No HN    & 43.2 & 1.5 & No  & 0.61,0.35,89.2    & 38  & 0.26 & ...  \\
               & 2009 Feb 12 & BP143C & No MK    & 43.2 & 1.5 & No  & 0.55,0.29,60.6    & 41  & 0.29 & ...  \\ \colrule
\end{tabular}}
\end{center}
{\scriptsize $a$: BR = Brewster, Washington, HN = Hancock, New Hampshire,
KP = Kitt Peak, Arizona, MK = Mauna Kea, Hawaii,
NL = North Liberty, Iowa, SC = Saint Croix, U.S.\\ Virgin Islands.}\\
{\scriptsize $b$: Whether or not the experiment recorded dual-circular polarization.}\\
{\scriptsize $c$: Numbers given are for the naturally weighted beam, and are the FWHMs of the major
and minor axes in mas, and the position angle of the major axis in degrees.
Position\\ angle is measured from north through east.}\\
{\scriptsize $d$: $I_{rms}$ and $P_{rms}$ are the rms noise levels in the total intensity and polarized intensity images,
respectively.}\\
\end{sidewaystable*}

VLBA experiment BE044 recorded single-epoch full-track observations of nine
TeV blazars during 2006 and 2007, including five of the six sources discussed in this paper
(Mrk~421, Mrk~501, H~1426+428, 1ES~1959+650, and 1ES~2344+514).
The other four TeV blazars observed during BE044 were more newly established TeV blazars
for which the VLBA data will be presented in a future paper.
Mrk~421, Mrk~501, 1ES~1959+650, and 1ES~2344+514 were all observed at 22~GHz
with dual-circular polarization (note that a VLBA polarization image of PKS~2155$-$304 has
already been published in P08).
The fainter source H~1426+428 was observed in total intensity
only, at a frequency of 8~GHz. All observations in BE044 were recorded at the then standard data rate of 128~Mbps. 
The 22~GHz observations were recorded prior to the recent 22~GHz sensitivity upgrade at the VLBA. 
Details of the five BE044 observations included in this paper are given in Table~\ref{imtab}.

VLBA experiment BP143 recorded three epochs each at 43 GHz on the five TeV blazars
Mrk~421, Mrk~501, 1ES~1959+650, PKS~2155$-$304, and 1ES~2344+514 during 2008 and 2009.
The observations were recorded at a high data rate of 512~Mbps to make sure of detecting fringes to
the comparatively fainter sources such as 1ES~1959+650 and 1ES~2344+514. All five sources were detected
at all epochs. Details of these 15 observations, which make up the bulk of the images for this paper,
are given in Table~\ref{imtab}.

We used the AIPS software package for calibration and fringe-fitting of the correlated
visibilities, and the visibilities were edited and final CLEAN images were produced
using the DIFMAP software package. 
In the section below, all images are shown using natural weighting (uvweight=0,$-$2~in DIFMAP).

For the polarization experiment BE044, calibration of
the polarization response of the feeds was done with LPCAL in AIPS.
The required electric vector position angle (EVPA) corrections were applied using CLCOR in AIPS, and
were determined from the observed EVPA of stable calibrator sources
compared with the EVPA recorded for these sources on
the VLA/VLBA Polarization Calibration Page\footnote{http://www.vla.nrao.edu/astro/calib/polar/},
interpolated to our epoch of observation.
A $\pm5\arcdeg$ uncertainty is estimated for the calibration of EVPA.
Final polarization images were made using FITSPLOT scripts\footnote{
personal.denison.edu/$\sim$homand/FITSplot.v202.tar.gz}.
No corrections for internal Faraday rotation could be derived for our single-frequency data; 
however, the rotation measures
in BL Lac objects at low frequencies are typically small (Zavala \& Taylor 2003; 2004).
Although these rotation measures can increase in the cores at higher frequencies 
(Jorstad et al. 2007; O'Sullivan \& Gabuzda 2009), 
the clear and physically significant
patterns seen in the transverse polarization structure of the jets (see $\S$~\ref{transverse})
argues against a large and variable internal Faraday rotation in these jets at 22~GHz. 

After final calibration of the visibilities, circular or elliptical Gaussian model components were fit to the visibilities
using the {\em modelfit} task in DIFMAP. Circular Gaussians were used by default, and provided satisfactory fits 
to the visibilities for
all epochs (as noted by the reduced chi-squared of the fit, and visual inspection of the residual map and visibilities).
Note that model fitting directly to the visibilities allows sub-beam
resolution to be obtained, and components can be clearly identified in the model fitting even when they
appear blended with the core component or with each other in the CLEAN images.
At some epochs for Mrk~421 and Mrk~501, a limb-brightened geometry offered an alternative model fit to circular Gaussians. 
In such a geometry, the core and each limb of the jet are fit by elliptical Gaussians, as was previously done for
Mrk~501~in P09. These fits are discussed further in the sections on those individual sources below.

The circular Gaussian models fit to all 23 observations in this paper are given in Table~\ref{mfittab}.
Note that flux values for closely spaced components may be inaccurate, since it is difficult for
DIFMAP to uniquely distribute the flux during model fitting.
The model component identification follows the naming scheme used in our previous papers (e.g., P08,
components are numbered starting at C1 from the outermost component inwards), and since the components
have been previously established to be quite slow in these sources, the identification of components across even
a time gap of several years turned out to be straightforward.
Brightness temperatures are also given in Table~\ref{mfittab} for all partially resolved core components.
The median core brightness temperature from Table~\ref{mfittab} is about $1\times10^{10}\rm~K$, similar to that
found in our previous papers (e.g., P08).
The alternate limb-brightened fits for some epochs for Mrk~421 and Mrk~501 are given in Table~\ref{limbfittab},
and may be compared with the similar fits given for Mrk~501 during 2005 by P09.

\begin{table*}[!t]
\caption{Circular Gaussian Models} 
\vspace{-0.15in}
\begin{center}
\label{mfittab}
{\scriptsize \begin{tabular}{l l c c c c r c c c} \colrule \colrule \\ [-9pt]
& & Frequency & & $S$ & $r$ &
\multicolumn{1}{c}{PA} & $a$ & & $T_{B}$ \\ 
\multicolumn{1}{c}{Source} & \multicolumn{1}{c}{Epoch} & (GHz) & Component & (mJy) & (mas) &
\multicolumn{1}{c}{(deg)} & (mas) & $\chi_{R}^{2}$ & ($10^{10}$ K) \\
\multicolumn{1}{c}{(1)} & \multicolumn{1}{c}{(2)} & (3) & (4) & (5) & (6) &
\multicolumn{1}{c}{(7)} & (8) & (9) & (10) \\ \colrule \\ [-5pt]
Markarian 421  & 2005 May 21 & 43.2 & Core & 176 & ...  & ...      & 0.08 & 0.66 & 1.9  \\
               &             &      & C7   & 19  & 0.53 & $-$26.4  & 0.46 &      &      \\
               & 2005 Jun 21 & 43.2 & Core & 123 & ...  & ...      & 0.00 & 0.84 & ...  \\
               &             &      & C7   & 9   & 0.40 & $-$39.2  & 0.21 &      &      \\
               & 2005 Sep 2  & 43.2 & Core & 183 & ...  & ...      & 0.06 & 0.90 & 3.4  \\
               &             &      & C7   & 16  & 0.47 & $-$14.6  & 0.33 &      &      \\
               & 2006 Oct 15 & 22.2 & Core & 222 & ...  & ...      & 0.07 & 0.76 & 11.6 \\
               &             &      & C7   & 34  & 0.29 & $-$29.3  & 0.16 &      &      \\
               &             &      & C6   & 26  & 0.75 & $-$31.7  & 0.40 &      &      \\
               &             &      & C5   & 14  & 2.11 & $-$24.4  & 0.98 &      &      \\
               & 2008 Dec 4  & 43.2 & Core & 201 & ...  & ...      & 0.07 & 0.74 & 2.8  \\
               &             &      & C8   & 9   & 0.21 & 1.6      & 0.03 &      &      \\
               &             &      & C7   & 11  & 0.42 & $-$42.5  & 0.16 &      &      \\
               &             &      & C6   & 9   & 0.91 & $-$32.0  & 0.46 &      &      \\
               & 2009 Jan 10 & 43.2 & Core & 205 & ...  & ...      & 0.07 & 0.68 & 2.8  \\ 
               &             &      & C8   & 23  & 0.27 & $-$24.0  & 0.37 &      &      \\
               &             &      & C6   & 11  & 1.35 & $-$23.1  & 0.56 &      &      \\
               & 2009 Feb 12 & 43.2 & Core & 167 & ...  & ...      & 0.12 & 0.66 & 0.8  \\
               &             &      & C8   & 10  & 0.21 & 23.6     & 0.02 &      &      \\
               &             &      & C7   & 16  & 0.48 & $-$33.0  & 0.33 &      &      \\
               &             &      & C6   & 9   & 1.02 & $-$27.8  & 0.62 &      &      \\
H~1426+428     & 2006 Oct 22 & 8.4  & Core & 18  & ...  & ...      & 0.17 & 0.81 & 1.2  \\
               &             &      & C1   & 3   & 1.86 & $-$11.9  & 2.15 &      &      \\
Markarian 501  & 2006 Oct 16 & 22.2 & Core & 549 & ...  & ...      & 0.17 & 1.12 & 4.9  \\
               &             &      & C4   & 149 & 0.72 & 154.0    & 0.61 &      &      \\
               &             &      & C3   & 95  & 2.40 & 149.3    & 1.27 &      &      \\
               & 2008 Dec 4  & 43.2 & Core & 297 & ...  & ...      & 0.18 & 1.49 & 0.6  \\
               &             &      & C5   & 75  & 0.14 & $-$170.1 & 0.09 &      &      \\
               &             &      & C4   & 115 & 0.68 & 142.1    & 1.38 &      &      \\
               & 2009 Jan 10 & 43.2 & Core & 239 & ...  & ...      & 0.18 & 1.41 & 0.5  \\
               &             &      & C5   & 130 & 0.11 & $-$174.5 & 0.12 &      &      \\
               &             &      & C4   & 104 & 0.60 & 136.8    & 1.17 &      &      \\
               & 2009 Feb 12 & 43.2 & Core & 261 & ...  & ...      & 0.20 & 0.94 & 0.4  \\
               &             &      & C5   & 57  & 0.08 & $-$171.2 & 0.00 &      &      \\
               &             &      & C4   & 67  & 0.58 & 149.8    & 0.67 &      &      \\
1ES~1959+650   & 2007 Jan 27 & 22.2 & Core & 96  & ...  & ...      & 0.12 & 0.78 & 1.7  \\
               &             &      & C2   & 48  & 0.53 & 150.6    & 0.49 &      &      \\
               & 2008 Dec 4  & 43.2 & Core & 60  & ...  & ...      & 0.06 & 0.69 & 1.1  \\
               &             &      & C3   & 17  & 0.20 & 158.8    & 0.20 &      &      \\
               &             &      & C2   & 30  & 0.60 & 144.8    & 0.37 &      &      \\
               & 2009 Jan 10 & 43.2 & Core & 54  & ...  & ...      & 0.08 & 0.61 & 0.6  \\
               &             &      & C3   & 16  & 0.17 & 155.0    & 0.22 &      &      \\
               &             &      & C2   & 24  & 0.67 & 143.2    & 0.58 &      &      \\
               & 2009 Feb 12 & 43.2 & Core & 58  & ...  & ...      & 0.09 & 0.70 & 0.5  \\
               &             &      & C3   & 14  & 0.19 & 152.7    & 0.17 &      &      \\
               &             &      & C2   & 25  & 0.56 & 144.8    & 0.41 &      &      \\
PKS~2155$-$304 & 2008 Dec 4  & 43.2 & Core & 124 & ...  & ...      & 0.09 & 0.67 & 1.1  \\
               &             &      & C3   & 52  & 0.23 & $-$124.5 & 0.14 &      &      \\
               &             &      & C2   & 29  & 0.39 & $-$177.1 & 0.30 &      &      \\
               &             &      & C1   & 11  & 1.43 & 163.7    & 1.30 &      &      \\
               & 2009 Jan 10 & 43.2 & Core & 70  & ...  & ...      & 0.07 & 0.62 & 1.0  \\
               &             &      & C3   & 87  & 0.21 & $-$126.8 & 0.31 &      &      \\
               &             &      & C1   & 12  & 1.46 & 160.0    & 1.28 &      &      \\
               & 2009 Feb 12 & 43.2 & Core & 94  & ...  & ...      & 0.09 & 0.69 & 0.8  \\
               &             &      & C3   & 60  & 0.16 & $-$120.3 & 0.25 &      &      \\
               &             &      & C2   & 53  & 0.53 & 172.7    & 0.43 &      &      \\
               &             &      & C1   & 8   & 1.50 & 155.4    & 1.04 &      &      \\ \colrule
\end{tabular}}
\end{center}
\end{table*}

\begin{table*}[!t]
\begin{center}
{\bf Table 2} (continued)\\ Circular Gaussian Models \\ [5pt]
{\scriptsize \begin{tabular}{l l c c c c r c c c} \colrule \colrule \\ [-9pt]
& & Frequency & & $S$ & $r$ &
\multicolumn{1}{c}{PA} & $a$ & & $T_{B}$ \\ 
\multicolumn{1}{c}{Source} & \multicolumn{1}{c}{Epoch} & (GHz) & Component & (mJy) & (mas) &
\multicolumn{1}{c}{(deg)} & (mas) & $\chi_{R}^{2}$ & ($10^{10}$ K) \\
\multicolumn{1}{c}{(1)} & \multicolumn{1}{c}{(2)} & (3) & (4) & (5) & (6) &
\multicolumn{1}{c}{(7)} & (8) & (9) & (10) \\ \colrule \\ [-5pt]
1ES~2344+514   & 2006 Oct 17 & 22.2 & Core & 107 & ...  & ...      & 0.05 & 0.65 & 11.1 \\
               &             &      & C3   & 5   & 0.77 & 142.5    & 0.17 &      &      \\
               &             &      & C2   & 3   & 1.60 & 136.8    & 0.38 &      &      \\
               &             &      & C1   & 5   & 3.98 & 143.4    & 2.38 &      &      \\
               & 2008 Dec 4  & 43.2 & Core & 25  & ...  & ...      & 0.00 & 0.80 & ...  \\
               &             &      & C4   & 17  & 0.09 & 121.0    & 0.22 &      &      \\
               &             &      & C3   & 3   & 0.64 & 151.1    & 0.32 &      &      \\
               & 2009 Jan 10 & 43.2 & Core & 41  & ...  & ...      & 0.11 & 0.66 & 0.2  \\
               &             &      & C3   & 3   & 0.82 & 135.3    & 0.14 &      &      \\
               & 2009 Feb 12 & 43.2 & Core & 39  & ...  & ...      & 0.08 & 0.64 & 0.4  \\ 
               &             &      & C4   & 10  & 0.21 & 127.8    & 0.17 &      &      \\
               &             &      & C3   & 4   & 0.76 & 136.0    & 0.41 &      &      \\ \colrule
\end{tabular}}
\end{center}
NOTES.--- Col.~(5): Flux density in millijanskys.
Col.~(6) and (7): $r$ and PA are the polar coordinates of the
center of the component relative to the presumed core.
Position angle is measured from north through east.
Col.~(8): $a$ is the Full Width at Half Maximum (FWHM) of the circular Gaussian
component.
Col.~(9): The reduced chi-squared of the model fit.
Col.~(10): The maximum source-frame brightness temperature of the circular Gaussian core component is given by
$T_{B}=1.22\times10^{12}\;\frac{S(1+z)}{a^{2}\nu^{2}}$~K,
where $S$ is the flux density of the Gaussian in Janskys,
$a$ is the FWHM of the Gaussian in mas,
$\nu$ is the observation frequency in GHz, and $z$ is the redshift.
Brightness temperature is given for core components whose best-fit size is not zero.
\end{table*}

\begin{table*}
\caption{Limb-brightened Elliptical Gaussian Models}
\vspace{-0.15in}
\begin{center}
\label{limbfittab}
{\small \begin{tabular}{l l c c c c c c c r} \colrule \colrule \\ [-9pt]
& & Frequency & & $S$ & $r$ &
PA & $a$ & & \multicolumn{1}{c}{$\phi$} \\ 
\multicolumn{1}{c}{Source} & \multicolumn{1}{c}{Epoch} & (GHz) & Component & (mJy) & (mas) &
(deg) & (mas) & $b/a$ & (deg) \\
\multicolumn{1}{c}{(1)} & \multicolumn{1}{c}{(2)} & (3) & (4) & (5) &
(6) & (7) & (8) & (9) & \multicolumn{1}{c}{(10)} \\ \colrule \\ [-5pt]
Markarian 421 & 2005 Sep 2  & 43.2 & Core         & 179 & ...  & ...      & 0.07 & 0.82 & $-$32.9 \\
              &             &      & Eastern limb & 12  & 0.56 & $-$7.2   & 0.82 & 0.00 & $-$22.9 \\
              &             &      & Western limb & 10  & 0.37 & $-$28.1  & 0.63 & 0.00 & $-$24.2 \\
              & 2008 Dec 4  & 43.2 & Core         & 193 & ...  & ...      & 0.09 & 0.67 & $-$43.8 \\
              &             &      & Eastern limb & 16  & 0.15 & 11.5     & 0.32 & 0.07 & $-$28.8 \\
              &             &      & Western limb & 21  & 0.50 & $-$40.8  & 0.82 & 0.23 & $-$36.0 \\
Markarian 501 & 2008 Dec 4  & 43.2 & Core         & 364 & ...  & ...      & 0.21 & 0.75 & $-$3.1  \\
              &             &      & Eastern limb & 77  & 0.42 & 115.7    & 1.39 & 0.33 & $-$18.9 \\
              &             &      & Western limb & 27  & 1.04 & 169.9    & 1.53 & 0.00 & $-$22.9 \\
              & 2009 Feb 12 & 43.2 & Core         & 295 & ...  & ...      & 0.17 & 0.89 & $-$19.8 \\
              &             &      & Eastern limb & 67  & 0.30 & 120.5    & 1.07 & 0.26 & $-$37.4 \\
              &             &      & Western limb & 26  & 0.69 & 171.5    & 0.63 & 0.00 & $-$36.7 \\ \colrule
\end{tabular}}
\end{center}
NOTES.--- Col.~(5): Flux density in millijanskys.
Col.~(6) and (7): Polar coordinates of the
center of the component relative to the core.
Position angle is measured from north through east.
Col.~(8), (9), and (10): FWHM of the major axis,
axial ratio, and position angle of the major axis.
\end{table*}

Polarization properties of the model components from the four polarization images are given in
Table~\ref{pfittab}, for all components where the peak polarized flux density of the component was above three
times the polarization rms noise level given in Table~\ref{imtab} (at this point, the error due to Ricean bias becomes
comparable to our calibration error, Wardle \& Kronberg [1974]). 
The polarization properties of the model components were determined by fitting to the Stokes
Q and U visibilities, with the positions and sizes of components fixed at their values from the total  
intensity model, as described by Homan et al. (2002). A global comparison of the polarization properties of
the TeV blazars to other blazar samples (e.g., MOJAVE) will be made in a forthcoming paper that also includes the
VLBI polarization properties of the more newly discovered TeV blazars.

\begin{table*}
\caption{Polarization Properties of Model Components}
\vspace{-0.15in}
\begin{center}
\label{pfittab}
{\small \begin{tabular}{l l c c c c r c} \colrule \colrule \\ [-9pt]
& & Frequency & & $P$ & &
\multicolumn{1}{c}{EVPA} & EVPA$-$JPA \\
\multicolumn{1}{c}{Source} & \multicolumn{1}{c}{Epoch} & (GHz) & Component & (mJy) & $m$ &
\multicolumn{1}{c}{(deg)} & (deg) \\ 
\multicolumn{1}{c}{(1)} & \multicolumn{1}{c}{(2)} & (3) & (4) & (5) & (6) & \multicolumn{1}{c}{(7)} &
(8) \\ \colrule \\ [-5pt]
Markarian 421 & 2006 Oct 15 & 22.2 & Core & 6.5  & 0.029 & 61.2    & 89.5 \\
              &             &      & C7   & 3.1  & 0.090 & $-$25.3 & 4.0  \\
              &             &      & C6   & 2.3  & 0.088 & 26.2    & 59.4 \\
Markarian 501 & 2006 Oct 16 & 22.2 & Core & 13.9 & 0.026 & $-$56.8 & 30.8 \\  
              &             &      & C4   & 8.4  & 0.057 & $-$11.1 & 14.9 \\
1ES~1959+650  & 2007 Jan 27 & 22.2 & Core & 3.7  & 0.039 & $-$59.4 & 30.0 \\  
              &             &      & C2   & 1.9  & 0.040 & $-$32.6 & 3.2  \\ 
1ES~2344+514  & 2006 Oct 17 & 22.2 & Core & 0.8  & 0.008 & 44.9    & 82.4 \\ \colrule
\end{tabular}}
\end{center}
NOTES.--- Col.~(5): 
Polarized flux density in millijanskys.
Col.~(6): Fractional polarization of component. An 10\% error in the flux scale
produces a 14\% error in fractional polarization.
Col.~(7): Electric Vector Position Angle of the polarized emission from the component.
The EVPA of the core may be affected by uncorrected Faraday rotation.
Col.~(8): Offset of the EVPA from the jet position angle. The jet position angle is defined
as the position angle from the component upstream, as in Lister \& Homan (2005).
\end{table*}

\section{Images and Model Fits for Individual Sources}
\label{results}

\subsection{Markarian 421}
VLBA images of Mrk~421 from the 43~GHz observations in 2005 (BP112) are shown in Figure~1.
Images from the 43~GHz observations in 2008 and 2009 (BP143) are shown in Figure~2.
The images in Figure~2 are of higher quality than those in Figure~1, because
of the greater amount of time spent on source during the 2008 and 2009 observations.
The linear scale on these images is 0.6~pc~mas$^{-1}$, for a redshift of $z=0.03$.
These images are the first 43~GHz images of this source that we have
obtained since the 1994 image published in Piner et al. (1999), and because of the higher data
rate used for these observations (512 Mbps), these images have significantly higher dynamic range.

\begin{figure*}
\begin{center}
\includegraphics[scale=0.75]{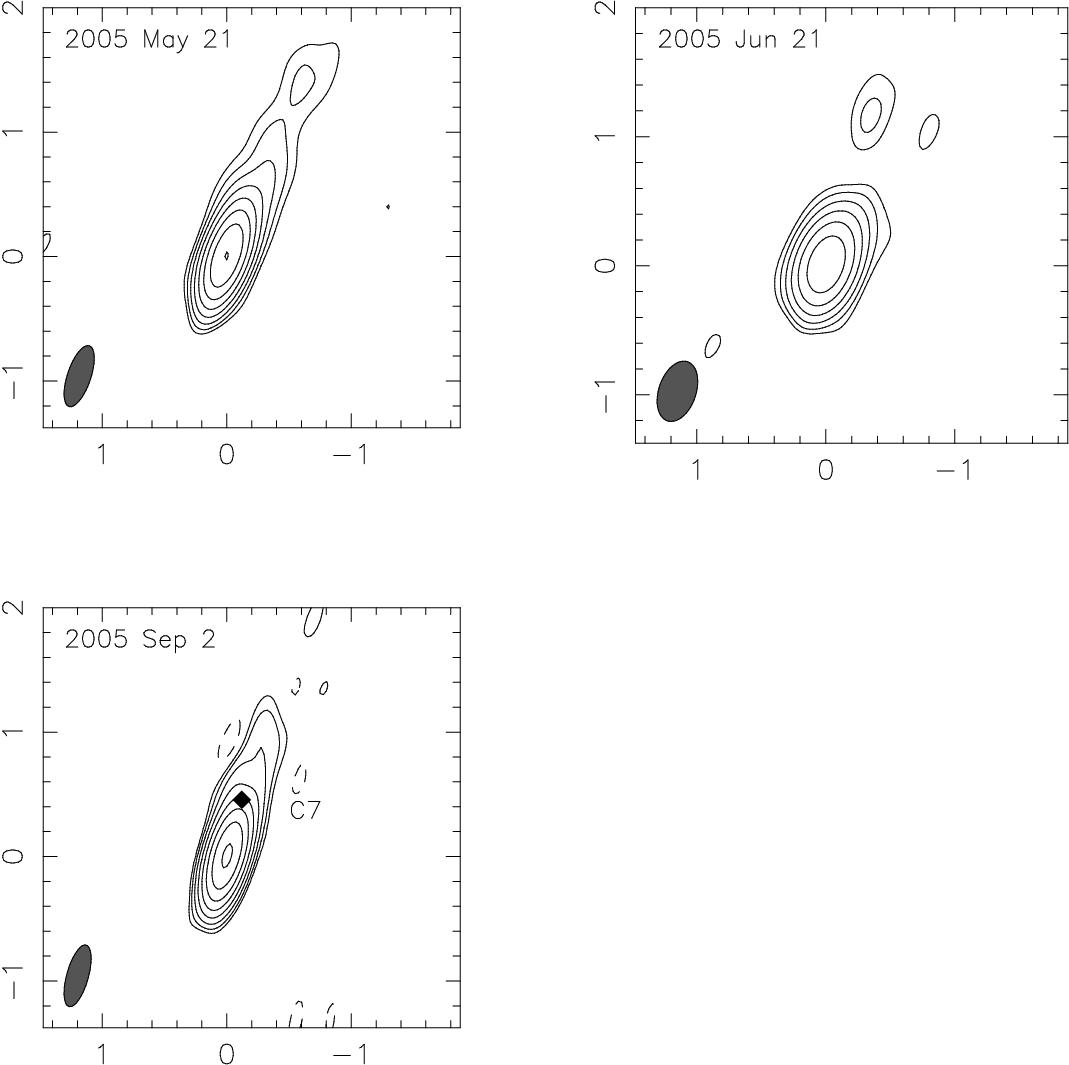}
\end{center}
\caption{VLBA images of Mrk~421 at 43 GHz during 2005.
The axes are labeled in milliarcseconds (mas).
The lowest contour is set to three times the rms noise level, and each
successive contour is a factor of two higher.
Numerical parameters of the images are given in Table~\ref{imtab}.
The position of the center of the circular Gaussian in the inner jet
that was fit to the visibilities is marked with a diamond at the last epoch.
Parameters of the circular Gaussian models are given in Table~\ref{mfittab}.}
\end{figure*}

\begin{figure*}
\begin{center}
\includegraphics[scale=0.75]{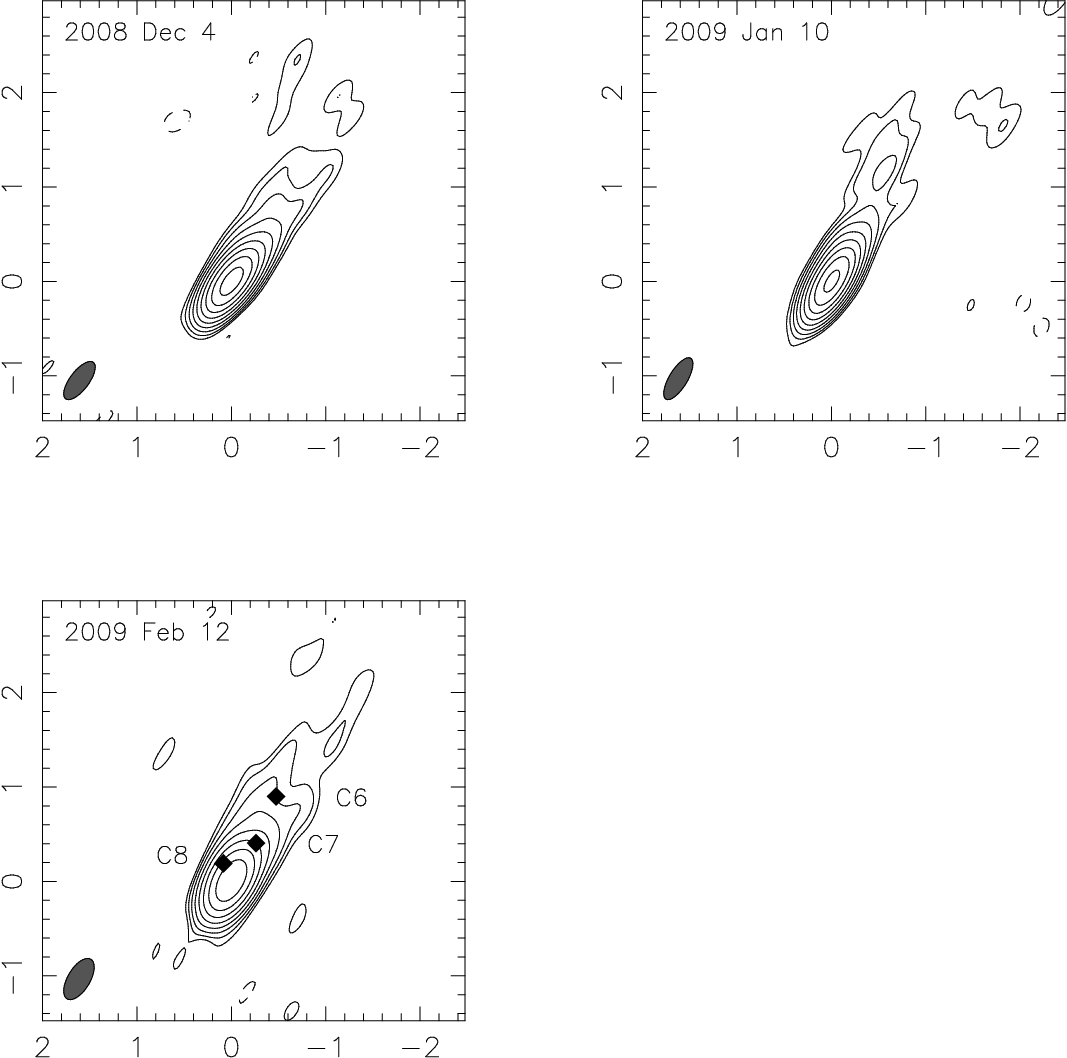}
\end{center}
\caption{VLBA images of Mrk~421 at 43 GHz during 2008 and 2009.
The axes are labeled in milliarcseconds (mas).
The lowest contour is set to three times the rms noise level, and each
successive contour is a factor of two higher.
Numerical parameters of the images are given in Table~\ref{imtab}.
The positions of the centers of the circular Gaussians in the inner jet
that were fit to the visibilities are marked with diamonds at the last epoch.
Parameters of the circular Gaussian models are given in Table~\ref{mfittab}.}
\end{figure*}

The images in Figure~2 show a limb-brightened jet in the inner milliarcsecond, extending
to the northwest from a $\sim200$~mJy core, in the same general direction as the 
structure visible at lower frequencies (Piner et al. 1999; Piner \& Edwards 2005).
A tentative detection of limb brightening in this source at 22~GHz was also reported
by Piner \& Edwards (2005), although it was not seen consistently at all distances from the core.
Giroletti et al. (2006) report a clear signature of limb brightening in 
their lower frequency 5~GHz VLBA images of Mrk~421 at several milliarcseconds from the core.
Here we confirm that this limb brightening extends into the inner milliarcsecond at 43~GHz.
Analysis of this limb-brightened structure is given
in $\S$~\ref{limb}.

Model components C6 and C7, which are familiar from earlier work (Piner \& Edwards 2005)
are detected in the 43~GHz datasets (see Table~\ref{mfittab}). In addition, a new model component, designated C8
\footnote{We do not assume that this is the same component designated C8 that was noted at a single
epoch in Piner et al. (1999)}, 
is detected about 0.2~mas almost due north of the core in the three highest quality
43~GHz observations (the ones from 2008 and 2009). 

A VLBA polarization image of Mrk~421 at 22~GHz is included in the montage of VLBA polarization
images in Figure~8. The polarization properties of the core and jet components C7 and C6
are given in Table~\ref{pfittab}. The core is 3\% polarized, and its 
EVPA is perpendicular to the jet; this is known to be one of two common states
for the core polarization of this source (Charlot et al. 2006, their Fig.~8).
In Piner \& Edwards (2005), the core EVPA had the opposite orientation, consistent
with the other predominant state described by Charlot et al. (2006).
Farther down the jet, at the location of C7, the EVPA has rotated to become parallel to the jet. 
We also detect a spine-sheath structure transverse to the jet in both EVPA and fractional polarization,
this structure is described further in $\S$~\ref{transversepol}.

\vspace*{0.25in}

\subsection{H~1426+428}
We show our single new image of H~1426+428 at 8~GHz from 2006 Oct 22~in Figure~3.
The linear resolution is about 2.3~pc~mas$^{-1}$, for a redshift of $z=0.129$. 
The source shows a 3~mJy jet component (identified with component C1 seen previously)
located about 2~mas northwest of the $\sim20$~mJy core.
The structure is quite similar to the four previous epochs for this source shown in P08.
An updated apparent speed based on the new measured position of C1 is given in $\S$~\ref{update}.

\begin{figure*}[!t]
\begin{center}
\includegraphics[scale=0.40]{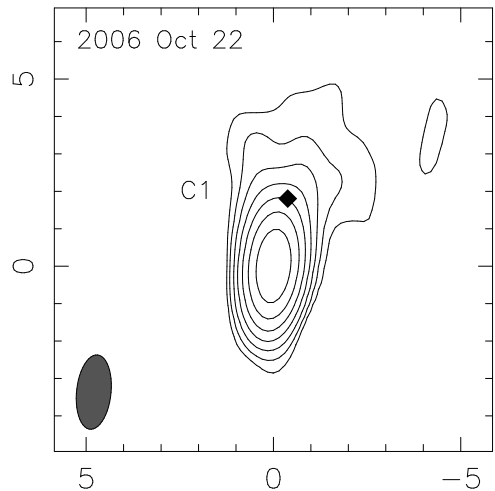}
\end{center}
\caption{VLBA image of H~1426+428 at 8 GHz on 2006 Oct 22.
The axes are labeled in milliarcseconds (mas).
The lowest contour is set to three times the rms noise level, and each
successive contour is a factor of two higher.
Numerical parameters of the image are given in Table~\ref{imtab}.
The position of the center of the circular Gaussian in the inner jet
that was fit to the visibilities is marked with a diamond.
Parameters of the circular Gaussian models are given in Table~\ref{mfittab}.}
\end{figure*}

\begin{figure*}
\begin{center}
\includegraphics[scale=0.75]{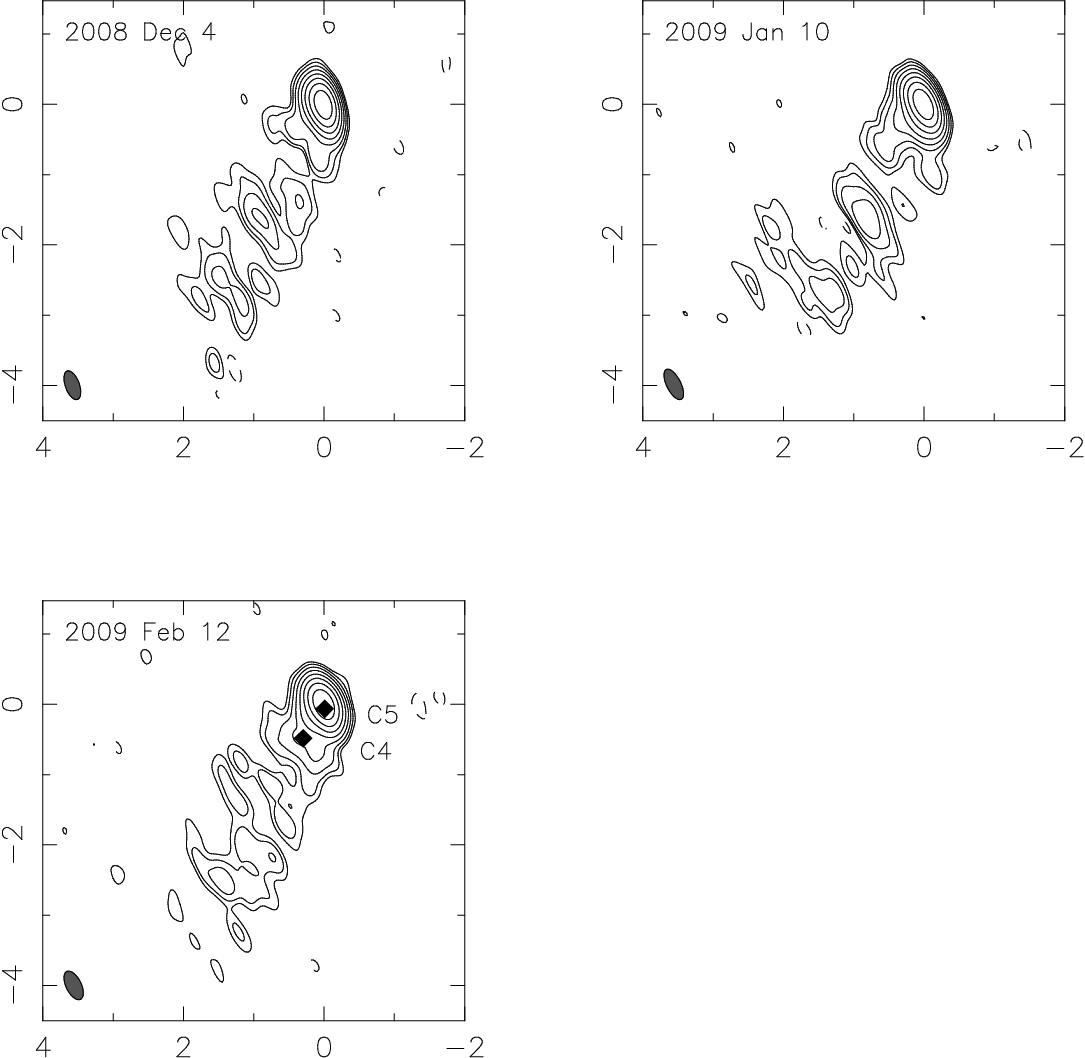}
\end{center}
\caption{VLBA images of Mrk~501 at 43 GHz during 2008 and 2009.
The axes are labeled in milliarcseconds (mas).
The lowest contour is set to three times the rms noise level, and each
successive contour is a factor of two higher.
Numerical parameters of the images are given in Table~\ref{imtab}.
The positions of the centers of the circular Gaussians in the inner jet
that were fit to the visibilities are marked with diamonds at the last epoch.
Parameters of the circular Gaussian models are given in Table~\ref{mfittab}.}
\end{figure*}

\subsection{Markarian 501}
Our VLBA images of Mrk~501 from the 43~GHz VLBA observations in 2008 and 2009 are shown in Figure~4.
The linear scale on these images is 0.7~pc~mas$^{-1}$, at a redshift of $z=0.034$.
The images shown a complex limb-brightened structure forming soon after the compact core region; this
limb-brightened structure has previously been discussed by us in the context of our 
43~GHz observations of this source from  2005 
(P09), and has also been described by other authors (e.g., Giroletti et al. 2008).

The orientation of the beam in the images in Figure~4 (with the major axis transverse to the jet
and the minor axis along the jet, necessitated by the scheduling of the five BP143 sources in one time block)
makes the limb-brightened structure not as well-resolved as in the images from 2005 shown by P09,
which had the orthogonal beam orientation.
Circular Gaussians provide an adequate fit to the inner jet at all three epochs, and models where
each limb of the jet is fit separately (as in P09)
only converge for two of the epochs (see Table~\ref{limbfittab}).
Therefore, we cannot confirm results relying on the motion of specific features in the limbs, such as
the tentative superluminal speed in the western limb from P09. 

The improved resolution along the jet allows us to detect a bright new model component, designated C5,
at about 0.1~mas from the core (at the same distance as features noted by Giroletti et al [2008] in
their 86~GHz image of this source). The position angle of this component, about $-170\arcdeg$, 
shows that the prominent bending of the jet, already well-measured at large scales,
continues in toward the core.

The VLBA polarization image at 22~GHz from 2006 is shown in the collection of images in Figure~8.
The polarization properties of the core and jet component C4 are given in Table~\ref{pfittab}.
As has been noted before (e.g., Pushkarev et al. 2005), Mrk~501 shows a spine-sheath structure in polarization,
with an EVPA parallel to the jet along the jet center, and an EVPA perpendicular to the jet along the edge.  
This structure
is particularly evident in our polarization image at about the location of component C4 (see Figure~8);
this structure is discussed further in $\S$~\ref{transversepol}.

\begin{figure*}
\begin{center}
\includegraphics[scale=0.75]{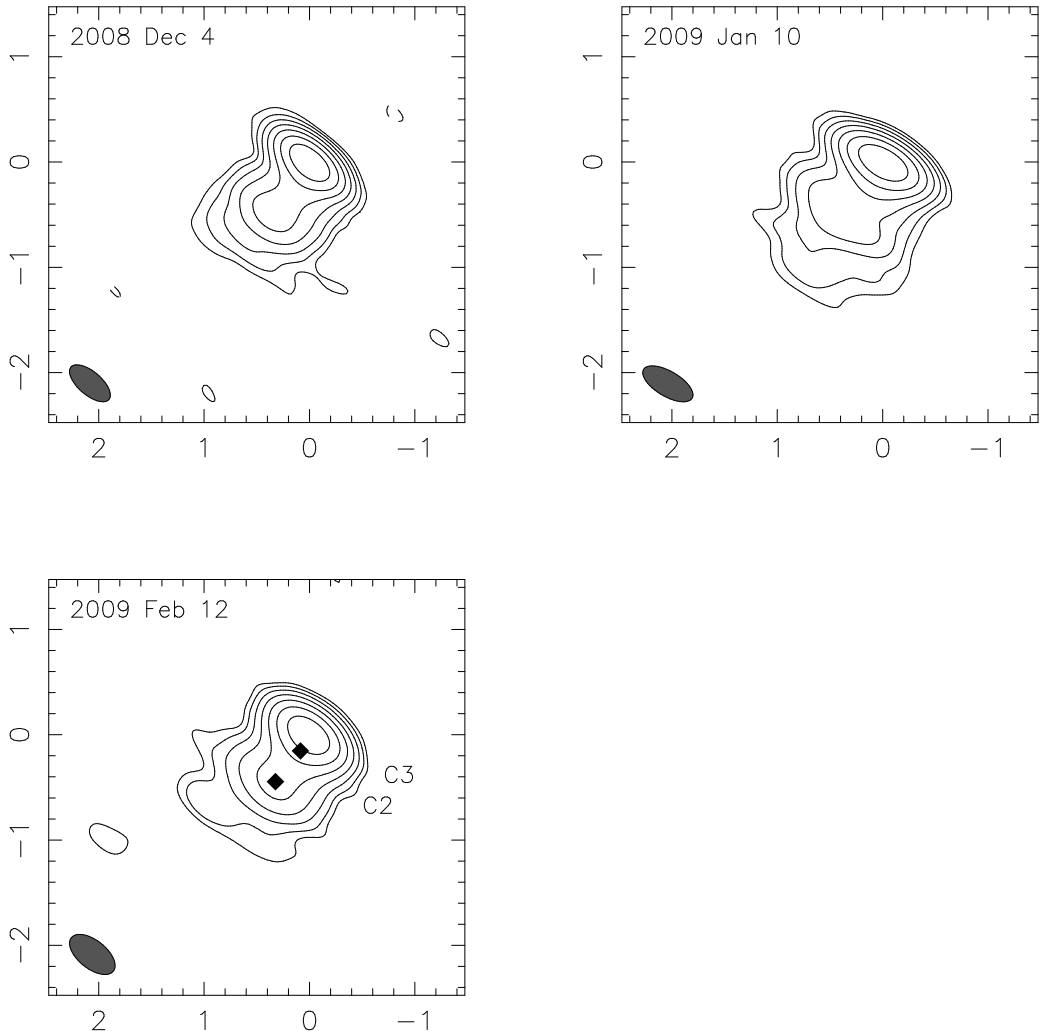}
\end{center}
\caption{VLBA images of 1ES~1959+650 at 43 GHz during 2008 and 2009.
The axes are labeled in milliarcseconds (mas).
The lowest contour is set to three times the rms noise level, and each
successive contour is a factor of two higher.
Numerical parameters of the images are given in Table~\ref{imtab}.
The positions of the centers of the circular Gaussians in the inner jet
that were fit to the visibilities are marked with diamonds at the last epoch.
Parameters of the circular Gaussian models are given in Table~\ref{mfittab}.}
\end{figure*}

\subsection{1ES~1959+650}
Figure~5 shows the 43~GHz images of 1ES~1959+650 from 2008 and 2009. At the distance of 1ES~1959+650 ($z=0.047$),
the linear scale of the images is 0.9~pc~mas$^{-1}$.
These are the first VLBA images of this source at 43~GHz, and thus they are the highest resolution images obtained for
this source to date.
At 43~GHz resolutions, 1ES~1959+650 consists of a compact core with a flux density of $\sim60$~mJy,
and a $\sim1$~mas jet extending to the southeast at a position angle of about $150\arcdeg$.
(See P08 for a discussion of the core identification in this source through spectral index mapping.)
Although the jet changes position angle abruptly at slightly larger scales (indeed, in lower-frequency 5~GHz VLBA images it 
is directed almost due north, see, e.g., Bondi et al. 2004), it is fairly straight at the scales shown here.
The jet structure is well-represented by two circular Gaussians, designated C2 and C3, with C3 being a
newly detected component about 0.2~mas from the core at a position angle of about $155\arcdeg$.

The VLBA polarization image at 22~GHz from 2007 is shown in the collection of images in Figure~8.
The polarization properties of the core and jet component C2 are given in Table~\ref{pfittab}.
The core and component C2 are both about 4\% polarized, and the EVPA is parallel to the
jet at the location of C2.
Although we do not detect any significant transverse jet structure in the total intensity images (Figure~5),
this source does display the same spine-sheath structure in both EVPA and fractional polarization that is also
displayed by Mrk~421 and Mrk~501 (easily noted around the location of C2 in Figure~8), that is
analyzed in detail in $\S$~\ref{transversepol}.

\subsection{PKS~2155$-$304}
Figure~6 shows the 43~GHz images of PKS~2155$-$304 from 2008 and 2009. At the distance of PKS~2155$-$304 ($z=0.116$),
the linear scale of the images is 2.1~pc~mas$^{-1}$.
These are the first VLBA images of this source at 43~GHz, and they are the highest resolution images published for
this source to date (by a factor of about three, since we had not previously observed this source at
frequencies higher than 15~GHz).

\begin{figure*}
\begin{center}
\includegraphics[scale=0.75]{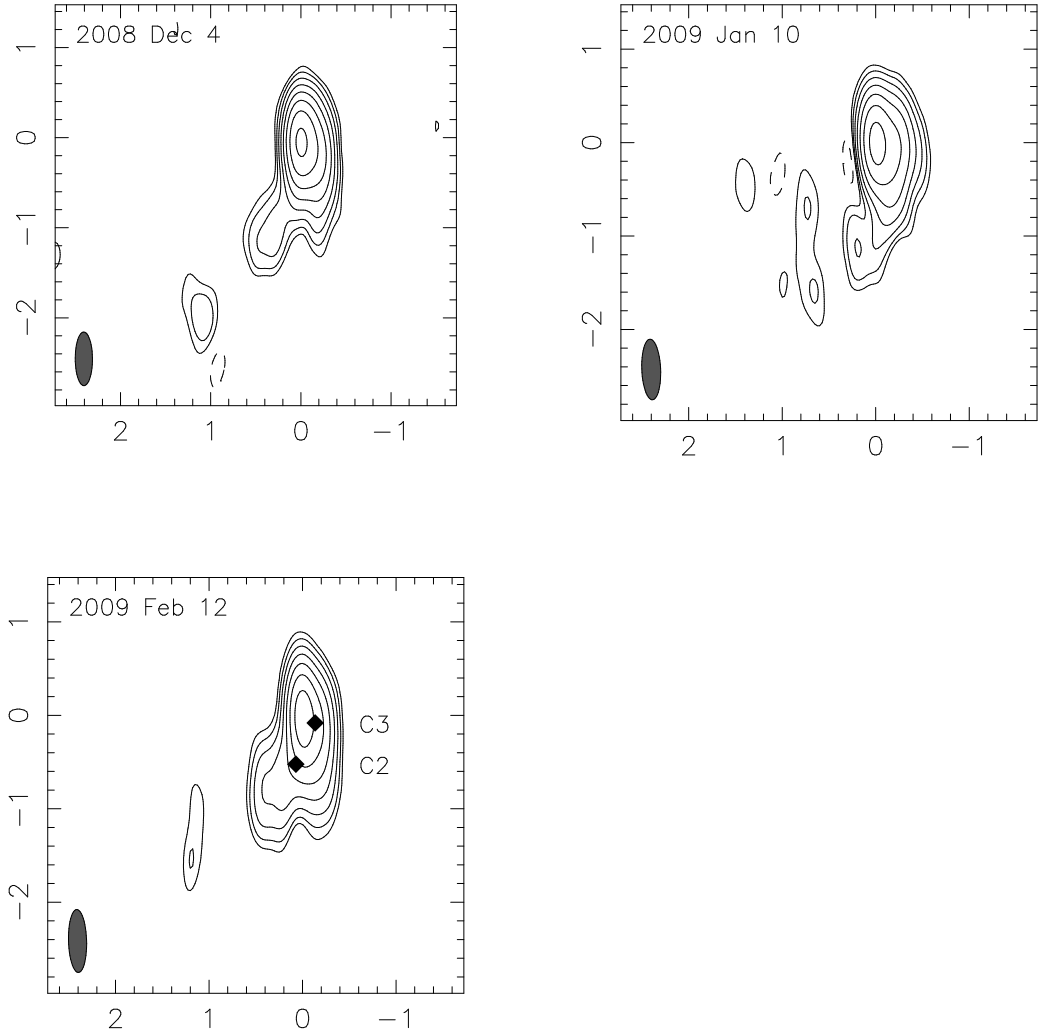}
\end{center}
\caption{VLBA images of PKS~2155$-$304 at 43 GHz during 2008 and 2009.
The axes are labeled in milliarcseconds (mas).
The lowest contour is set to three times the rms noise level, and each
successive contour is a factor of two higher.
Numerical parameters of the images are given in Table~\ref{imtab}.
The positions of the centers of the circular Gaussians in the inner jet
that were fit to the visibilities are marked with diamonds at the last epoch.
Parameters of the circular Gaussian models are given in Table~\ref{mfittab}.}
\end{figure*}

These new images reveal that the jet of PKS~2155$-$304 is strongly bent within the inner milliarcsecond.
(This can be particularly seen by following the contours on the 2009 Jan 10 image in Figure~6.)
The jet begins to the southwest at a position angle $-125\arcdeg$ before quickly bending due south of
the core and then approaching the 15~GHz position angle of $160\arcdeg$ by about 1~mas from the core.
This makes PKS~2155$-$304 the third of the six sources discussed in this paper to show extreme ($>\sim75\arcdeg$) jet bending
on parsec-scales (the other two being Mrk~501 and 1ES~1959+650). 
That 50\% of these sources show such large misalignments is a strong confirmation that the TeV
blazars as a class have their jets oriented quite close to the line of sight, where small bends can be amplified
by large projection effects.

The model fitting of Gaussian components to the jet follows the structure described above. The core component
is about 100~mJy at 43~GHz, and component C1, known from earlier studies, continues to be detected
about 1.5~mas from the core at a position angle of about $160\arcdeg$. In addition, these higher resolution
datasets reveal two new components, designated C2 and C3, at distances of about 0.5 and 0.2~mas from the core,
and position angles of about $180\arcdeg$ and $-125\arcdeg$, respectively.

A polarization observation of this source was not made in experiment BE044 (and thus it is not included
in the collection of polarization images in Figure~8), 
because the polarization structure of this source has been previously analyzed in P08.

\subsection{1ES~2344+514}
Figure~7 shows the 43~GHz images of 1ES~2344+514 from 2008 and 2009. At the distance of 1ES~2344+514 ($z=0.044$),
the linear scale of the images is 0.9~pc~mas$^{-1}$.
These are the first VLBA images of this source at 43~GHz, and thus they are the highest resolution images obtained for
this source to date (by a factor of about three, since we had not previously observed this source at
frequencies higher than 15~GHz).
This source is at the VLBA's detection limit at 43~GHz, with a data rate of 512 Mbps, and has a total compact flux 
of only about $\sim50$~mJy at this frequency.
The images in Figure~7 show a compact core with a flux of about $\sim40$~mJy, and a jet extending to the
southeast, in the direction of the jet seen at 15~GHz (Piner \& Edwards 2004). The jet in this source is fairly straight
at these scales. The model fitting at 43~GHz shows jet component C3 at about 0.7~mas from the core, 
and a new component, C4, at 0.1 to 0.2~mas from the core.

\begin{figure*}
\begin{center}
\includegraphics[scale=0.75]{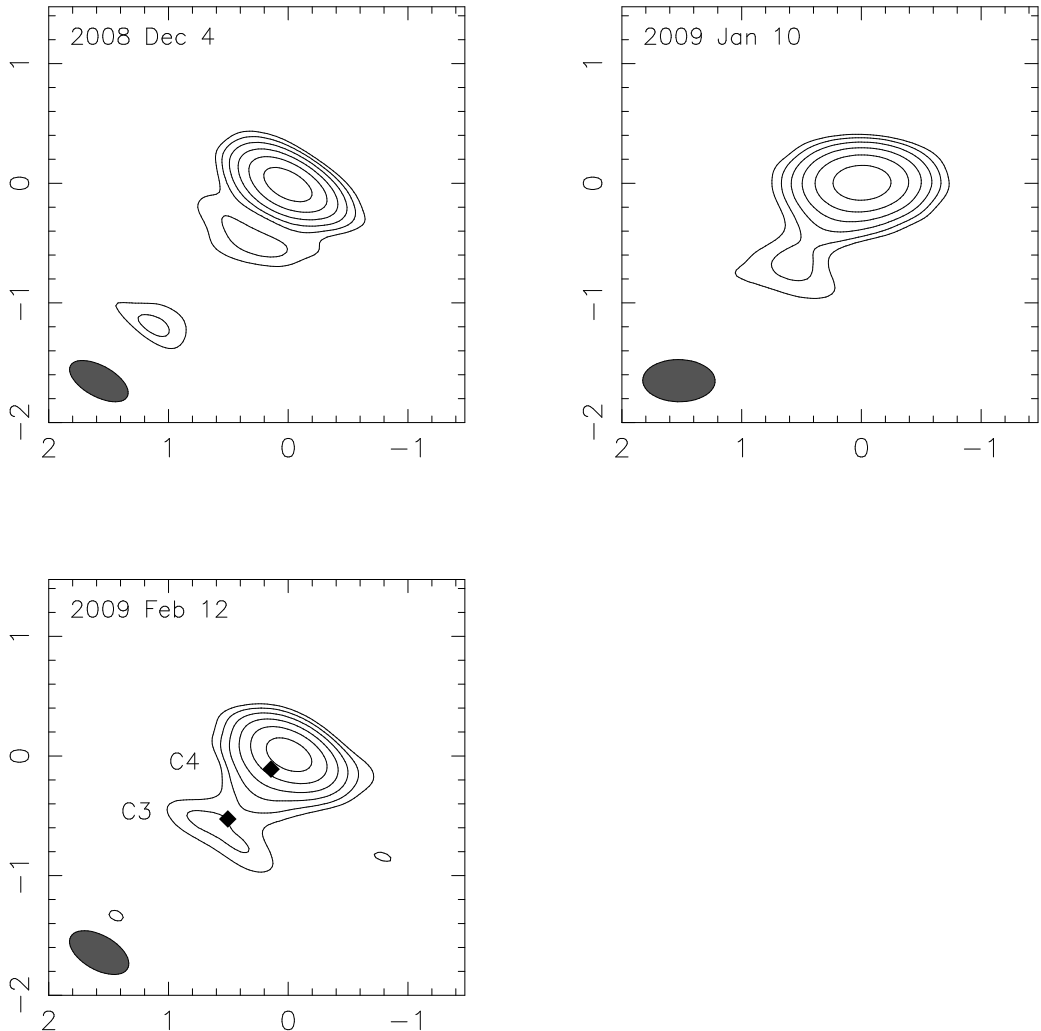}
\end{center}
\caption{VLBA images of 1ES~2344+514 at 43 GHz during 2008 and 2009.
The axes are labeled in milliarcseconds (mas).
The lowest contour is set to three times the rms noise level, and each
successive contour is a factor of two higher.
Numerical parameters of the images are given in Table~\ref{imtab}.
The positions of the centers of the circular Gaussians in the inner jet
that were fit to the visibilities are marked with diamonds at the last epoch.
Parameters of the circular Gaussian models are given in Table~\ref{mfittab}.}
\end{figure*}

The VLBA polarization image at 22~GHz from 2006 is shown in the collection of images in Figure~8.
The polarization properties of the core are given in Table~\ref{pfittab}.
The source is not highly polarized, and polarization is only significantly detected in the region of the core.
The core is about 1\% polarized (in agreement with the value found at lower frequencies by Bondi et al. 2004), 
and has an EVPA nearly orthogonal to the jet. 

\begin{figure*}
\begin{center}
\includegraphics[scale=0.84]{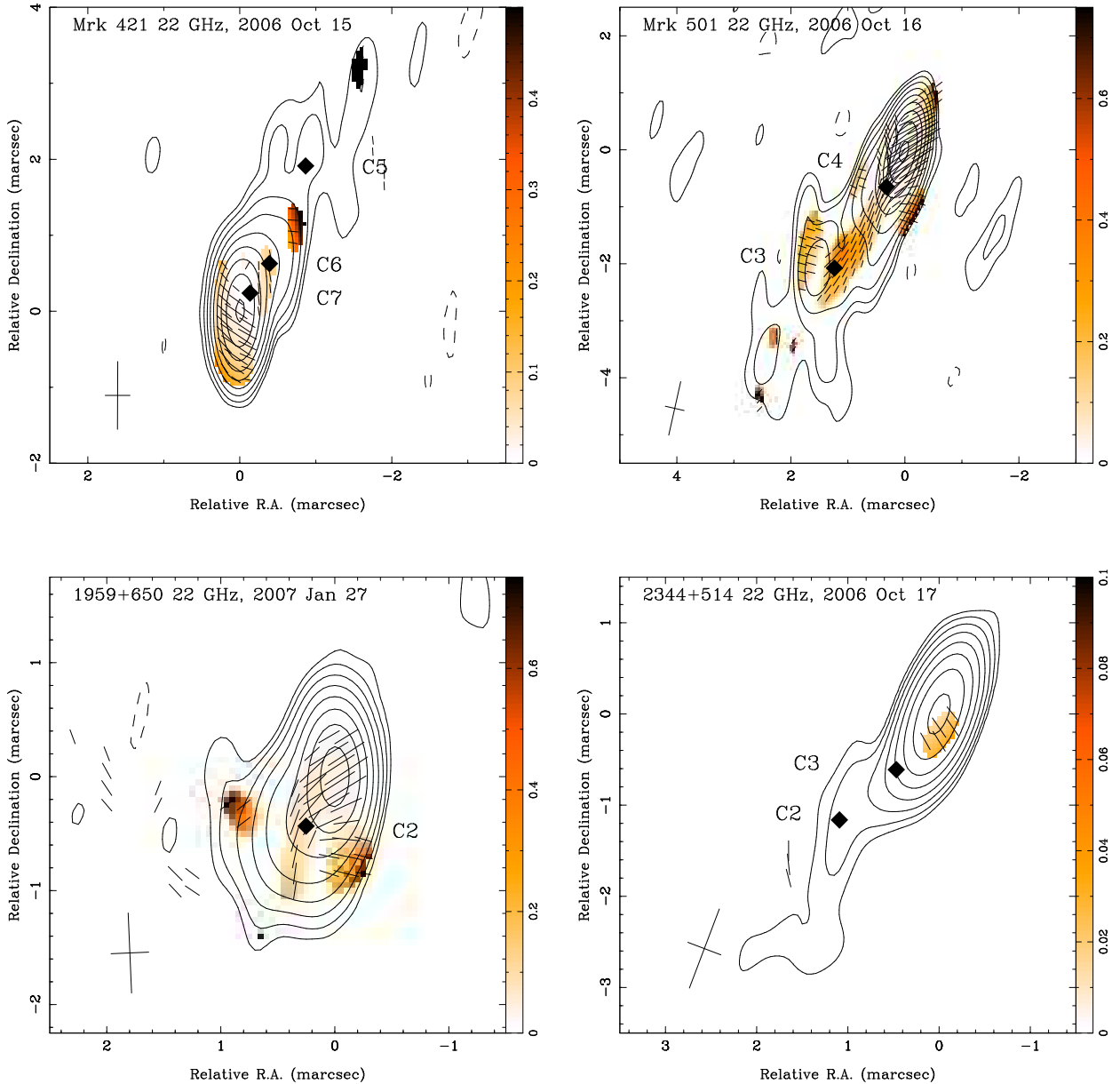}
\end{center}
\caption{
VLBA polarization images at 22 GHz. Contours show total intensity; lowest contours are
set to three times the rms noise, and each successive contour is a factor of two higher.
Tick marks show the direction of the EVPA and the magnitude of the
polarized flux (the scale is 75~mas~Jy$^{-1}$ for Mrk~421 and Mrk~501, and 150~mas~Jy$^{-1}$ for 
1ES~1959+650 and 1ES~2344+514). The color scale shows fractional polarization.
Tick marks and fractional polarization are drawn where the polarized flux is greater than 
six times the rms noise in the polarization image.
Numerical parameters of the images are given in Table~\ref{imtab}.
Locations of circular Gaussian components from model fitting are marked by diamonds.}
\end{figure*}

\section{Discussion}
\label{discussion}

\subsection{Transverse Structures}
\label{transverse}
When observed at high-resolution, three of these sources show structure transverse to the jet axis.
The limb-brightened structure seen in total intensity images of Mrk~501 at 43~GHz has already
been reported on in P09 (and at other frequencies by Giroletti et al. [2008, 2004b]).
Here we describe a similar limb-brightened structure in Mrk~421, and transverse polarization structures
in Mrk~421, Mrk~501, and 1ES~1959+650.

\subsubsection{Limb-Brightening in Mrk~421}
\label{limb}
Limb-brightening is evident in the jet of Mrk~421~in the 43~GHz VLBA images shown in Figure~2.
In Figure~9, we show an image of Mrk~421 from Figure~2 processed to highlight this limb brightening. In this figure,
the color-scale image is a total intensity image of the source with the core emission subtracted,
and super-resolved by a factor of two in the transverse direction.
With the core emission removed, the limb-brightened jet emission is
prominent, and reminiscent of that seen in Mrk~501 (P09; Giroletti et al. 2008, 2004b),
and in the nearby radio galaxy M87 (Kovalev et al. 2007; Ly et al. 2007).
Note that the intensity peaks in the two limbs correspond to 
components C8 (eastern limb) and C7 (western limb) from the Gaussian model fitting.

\begin{figure*}
\begin{center}
\includegraphics[scale=0.45]{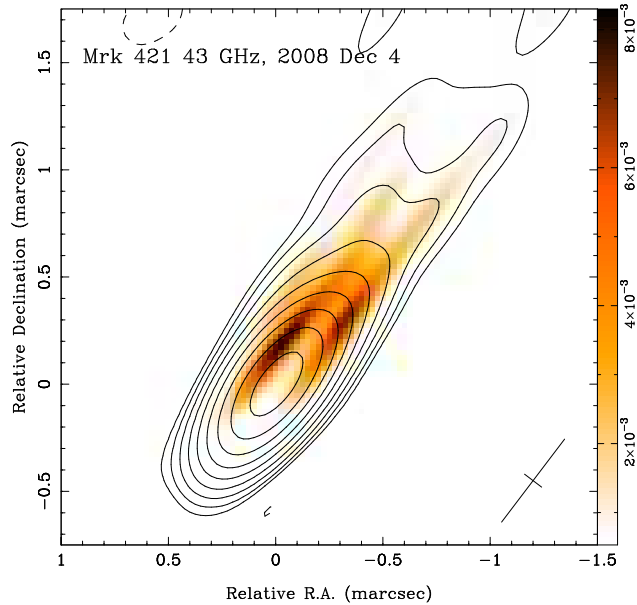}
\end{center}
\caption{Limb brightening in Mrk~421. The contour image is
the same as that shown for the corresponding date in Figure~2.
The color image is the same total intensity image 
with the CLEAN components representing the core subtracted, and
super-resolved by a factor of two transverse to the jet.
The super-resolved beam is shown at the bottom right.
With the contaminating core emission removed, the limb-brightened jet emission is
very prominent.}
\end{figure*}

\begin{figure*}
\begin{center}
\includegraphics[scale=0.50]{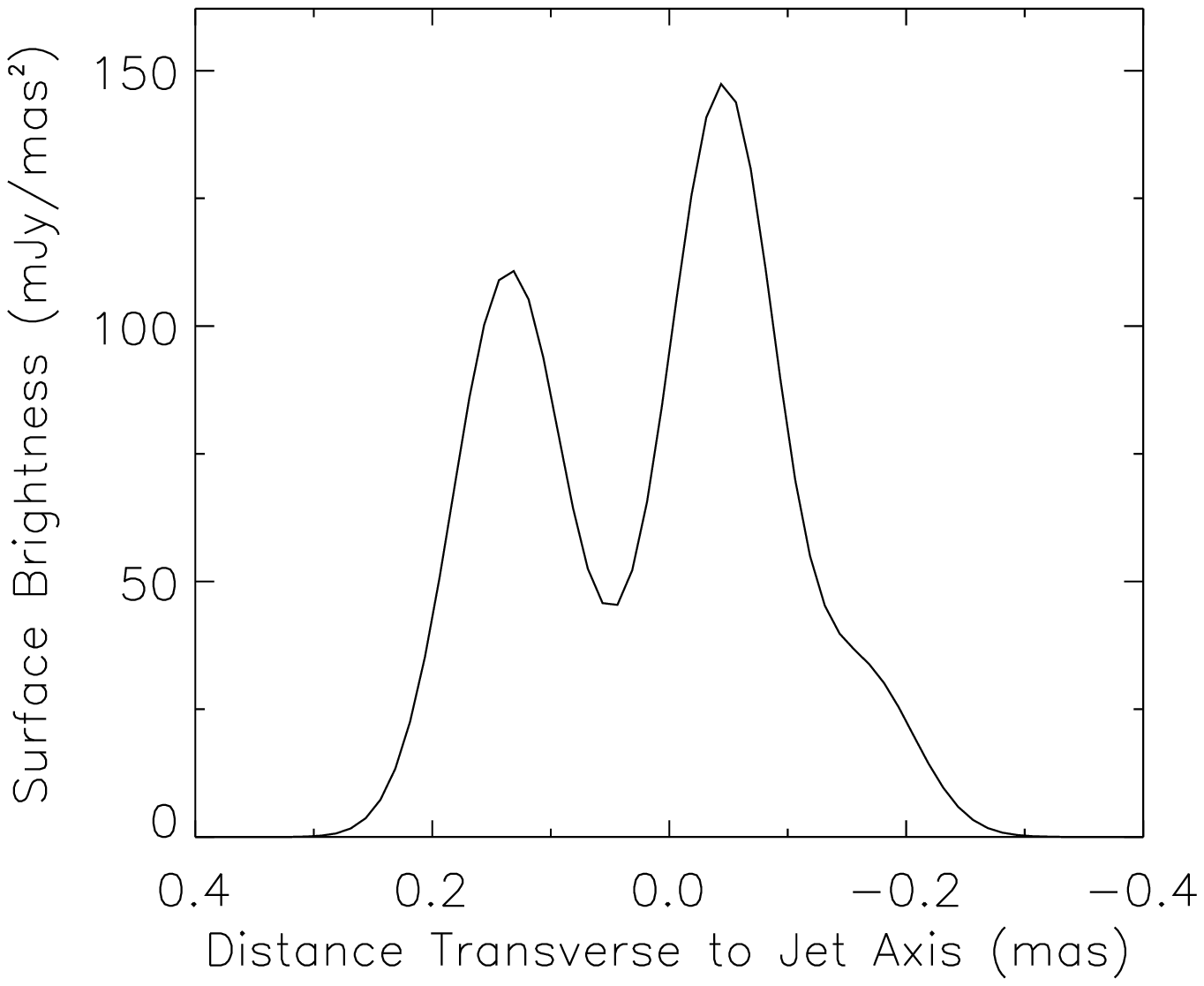}
\end{center}
\caption{Transverse brightness across the jet of Mrk~421 from the
color image in Figure~9, at a distance of 0.25~mas (0.15~pc projected) from the core.
The CLEAN components representing the core have been subtracted to avoid contamination
from core emission.}
\end{figure*}

Figure~10 shows a transverse slice across the color image from Figure~9, at 0.25~mas 
(0.15~pc projected) from the core.
The surface brightness ratio between the limbs and the center is greater than about 3:1
(this number is a lower limit because of beam convolution effects).
This slice resembles the similar slice shown for Mrk~501 by P09. 
Transverse slices that show limb-brightened structure in Mrk~421 have been previously shown
by Giroletti et al. (2006) at distances down to 2~mas from the core, and by Piner \& Edwards (2005)
at distances down to 3~mas from the core (but in Piner \& Edwards [2005], limb-brightening was
ambiguous at other distances).
Here we confirm that a clear limb-brightened structure extends down to at least 0.2~mas from
the core, or an order of magnitude closer than the structure reported by Giroletti et al. (2006). 

Limb-brightening in blazar jets may be caused by
either a higher Doppler factor or a higher synchrotron emissivity in the jet limbs
relative to the jet centers.
For example, Giroletti et al. (2004b, 2006) have interpreted the limb-brightened structures of both
Mrk~501 and Mrk~421~in terms of a fast spine and slower sheath. After starting out at a few degrees
to the line-of-sight (in the gamma-ray region), these jets may bend away from the line-of-sight,
allowing the slower sheath to acquire the larger Doppler factor. 
Specifically, Giroletti et al. (2006) consider $\Gamma_{\rm spine}\sim10$,
$\Gamma_{\rm sheath}\sim2$ and $\theta=18\arcdeg$ for the VLBI jet of Mrk~421,
giving $\delta_{\rm spine}\sim2$ and $\delta_{\rm sheath}\sim3$, and a sheath to spine
brightness ratio of about 3:1, consistent with what we observe in Figure~10. 
Giroletti et al. (2004b) invoke a similar bending geometry for Mrk~501.

Alternatively, the jet may remain at a constant (small) viewing angle, while a greater amount
of synchrotron radio emission is produced in the layer, causing the limb brightening.
For example, Sahayanathan (2009) explains the limb-brightening in Mrk~501 by shear
acceleration of electrons at the jet boundary.
Zakamska et al. (2008) find a solution for relativistic jet flow that involves
a pileup of material along the jet boundary, producing a limb-brightened structure
in synchrotron emission. Limb-brightening has also been interpreted in terms of
the Kelvin-Helmholtz instability acting at the jet boundary; see the discussion of
these models in the context of the M87 jet by Kovalev et al. (2007).
Transverse velocity structures may also naturally arise in these models as well.  

It is significant that extremely similar transverse structures are now seen in the two
closest and brightest TeV blazars
(Mrk~421 and Mrk~501). If this becomes established as a common property of the TeV blazars, 
then it would be difficult to
interpret all of these cases in terms of geometrical models that require precise
amounts of jet bending (Giroletti et al. 2004b, 2006), 
and models where such emission structures arise independent of a specific bending geometry would be
favored.

\begin{figure*}
\begin{center}
\includegraphics[scale=0.60]{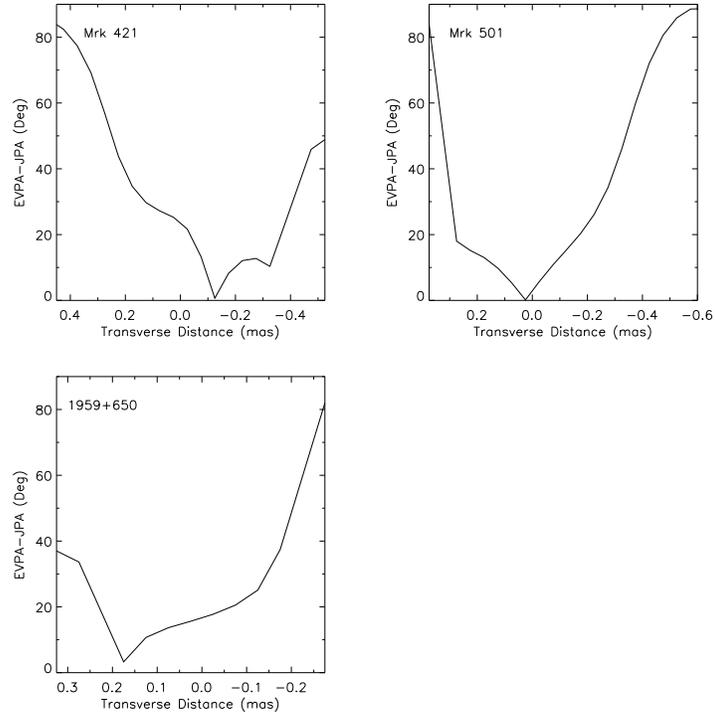}
\end{center}
\caption{EVPA$-$JPA (Electric Vector Position Angle compared to the
Jet Position Angle) versus transverse distance from the jet axis in Mrk~421, Mrk~501, and 1ES~1959+650
at a separation of 0.7 mas from the core. Upper left: Mrk~421.
Upper right: Mrk~501. Lower left: 1ES~1959+650}
\end{figure*}

\begin{figure*}
\begin{center}
\includegraphics[scale=0.60]{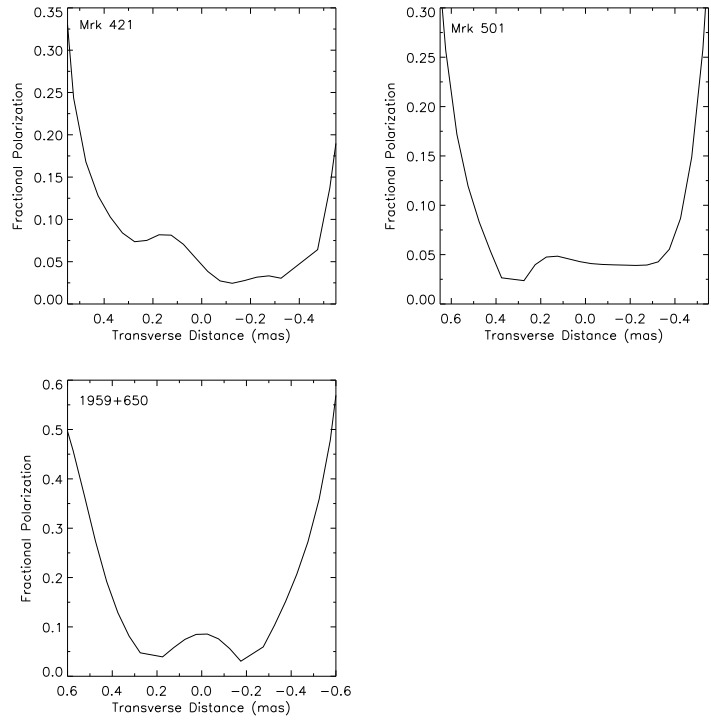}
\end{center}
\caption{Fractional polarization
versus transverse distance from the jet axis in Mrk~421, Mrk~501, and 1ES~1959+650
at a separation of 0.7 mas from the core. Upper left: Mrk~421.
Upper right: Mrk~501. Lower left: 1ES~1959+650}
\end{figure*}

\subsubsection{Transverse Polarization Structures}
\label{transversepol}
In this subsection, we discuss the transverse structures that are seen in the linear polarization
images of Mrk~421, Mrk~501, and 1ES~1959+650 at 22~GHz. Those polarization images are displayed in Figure~8. 
We first note that all of the innermost jet components detected with significant polarization
(C7 in Mrk~421, C4 in Mrk~501, and C2 in 1ES~1959+650, see Table~\ref{pfittab}) have their EVPA essentially
parallel to the jet direction, as is also common in the BL Lac objects 
present in the MOJAVE survey, which are predominantly LBLs (Lister \& Homan 2005).
Inspection of those jet regions in Figure~8 also shows that transverse structure in the linear polarization
is evident in those same regions; this transverse structure in all three of these sources shows what is commonly referred to
as a spine-sheath structure, with a parallel EVPA along the jet axis, and an orthogonal EVPA at
the jet edges, along with an increase in fractional polarization at the jet edges.

This spine-sheath polarization structure has already been well established by other authors for
Mrk~501 (Gabuzda et al. 2004; Pushkarev et al. 2005; Giroletti et al. 2008;
Croke et al. 2010). Here we confirm this structure in Mrk~501 at 22~GHz, and
we report the detection of similar transverse polarization structures in both Mrk~421 and 1ES~1959+650.
Figure~11 shows the distribution of EVPA$-$JPA (Electric Vector Position Angle compared to the
Jet Position Angle) versus transverse distance from the jet axis in Mrk~421, Mrk~501, and 1ES~1959+650
at a separation of 0.7 mas from the core in each source.
(A separation of 0.7 mas was chosen in order to be significantly far from the core, but still
in a region of high SNR.)
All three plots are very similar, with the EVPA$-$JPA approaching $0\arcdeg$ in the center of the jet,
and $90\arcdeg$ at at least one of the edges of the jet. The transverse distances chosen 
in Figure~11 show the maximum extent of the EVPA rotation in each jet.

Figure~12 shows the distribution of fractional polarization along
the same three slices as in Figure~11. The fractional polarization behavior is also similar for all
three sources, with fractional polarization increasing rapidly near the jet edges. 
This fractional polarization effect was also seen for Mrk~501 by Pushkarev et al. (2005) (their Figure 11).
Note that the EVPA rotation shown in Figure~11 occurs closer to the jet axis than the fractional polarization
increase shown in Figure~12, so that Figure~12 shows a greater transverse distance for each source.
In both Figure~11 and Figure~12, all plotted points have total and polarized intensities exceeding
three times the corresponding noise levels. Statistical errors in the plotted quantities are smaller 
near the jet axis (higher SNR), and rise to up to $\sim10\arcdeg$ in EVPA (Figure~11) and $\sim0.15$~in
fractional polarization (Figure~12) near the lower SNR jet edges. 

The theory of such transverse polarization structures in the context of the intrinsic magnetic
field geometry has been discussed by, e.g., 
Lyutikov et al. (2005) and Zakamska et al. (2008).
(Rotation measure gradients in some blazar jets [e.g., Asada et al. 2002; Gabuzda et al. 2004; G{\'o}mez et al. 2008]
argue in favor of such structures being due to intrinsic magnetic field 
geometry, rather than a series of shocks interacting with a surrounding medium.) 
Lyutikov et al. (2005) argue that transverse EVPA distributions like those in Figure~11
are consistent with a large-scale helical magnetic field in a resolved cylindrical jet.
They conclude that these structures provide evidence that ``the jet emissivity is confined to a narrow
region of radii at the periphery of the jet, where the toroidal magnetic field is of order of the
poloidal field in the jet frame." Such an emissivity distribution would also be consistent with
the limb-brightening observed in Mrk~421 and Mrk~501. Also, for certain combinations of viewing
angle and pitch angle in the Lyutikov et al. (2005) model, 
fractional polarization increases toward the jet edges, as seen in our fractional polarization distributions
in Figure~12 (see, for comparison, Figure~9 of Lyutikov et al. [2005]).
Note also that in the Lyutikov et al. (2005) model asymmetries in the fractional polarization
distribution have the potential to reveal the spin direction of the central engine. Since our fractional
polarization distributions are fairly symmetric, we do not attempt this here.

Zakamska et al. (2008) have also calculated the expected observed transverse 
intensity, EVPA, and fractional polarization
distributions in a model parabolic jet (i.e., as opposed to cylindrical or conical)
with a toroidal magnetic field, for both limb-brightened
and center-brightened jets. In their solution yielding an edge-brightened intensity distribution
(see their Figure~4), 
projection effects result in the characteristic spine-sheath EVPA distribution even when the 
magnetic field is purely toroidal (rather than helical).
However, they find that the polarization fraction is expected to rise toward the jet edges only
in their center-brightened model. This contradicts with our results from Figures~10 and 12, which show
that {\em both} the intensity and the fractional polarization increase at the jet edges.

Regardless of the precise magnetic field configuration responsible for these distributions, we stress, 
as we did in the previous section on limb-brightening, that
it is significant that the three TeV blazars that are both close enough and bright enough
to have their transverse polarization structure resolved by the VLBA all show such similar structures
(Figures~11 and 12), and that they therefore likely share a common physical mechanism that
establishes these structures.

\subsection{Kinematics}
\label{kinematics}
\subsubsection{Updated Kinematics of Previously Detected Components}
\label{update}
In this section, we present the latest results for the apparent jet speeds in these six TeV
blazars, based on the new epochs of data. Figure~13 shows separation versus time
plots for the model-fit Gaussian components in
all six blazars discussed in this paper, including epochs from the beginning of our monitoring
through the present paper. Error bars on model component positions in Figure~13 are calculated
differently than in earlier papers in this series (e.g., P08), which used a fixed
fraction of the beam for each component. Because the required fraction of a beam depended 
on component flux density, size, etc., that method became unwieldy with larger numbers of components.
The errors are now computed by the method described by Homan et al. (2001), and are computed 
from the scatter of the data points about the best linear fit (we use first-order fits rather than
the second-order fits described by Homan et al. [2001]).
When a component had three or fewer observations, then the error bar size for that component was set equal to
the error bar size of the next component out (or in if there was no component farther out).

\begin{figure*}[!t]
\begin{center}
\includegraphics[angle=90,scale=0.45]{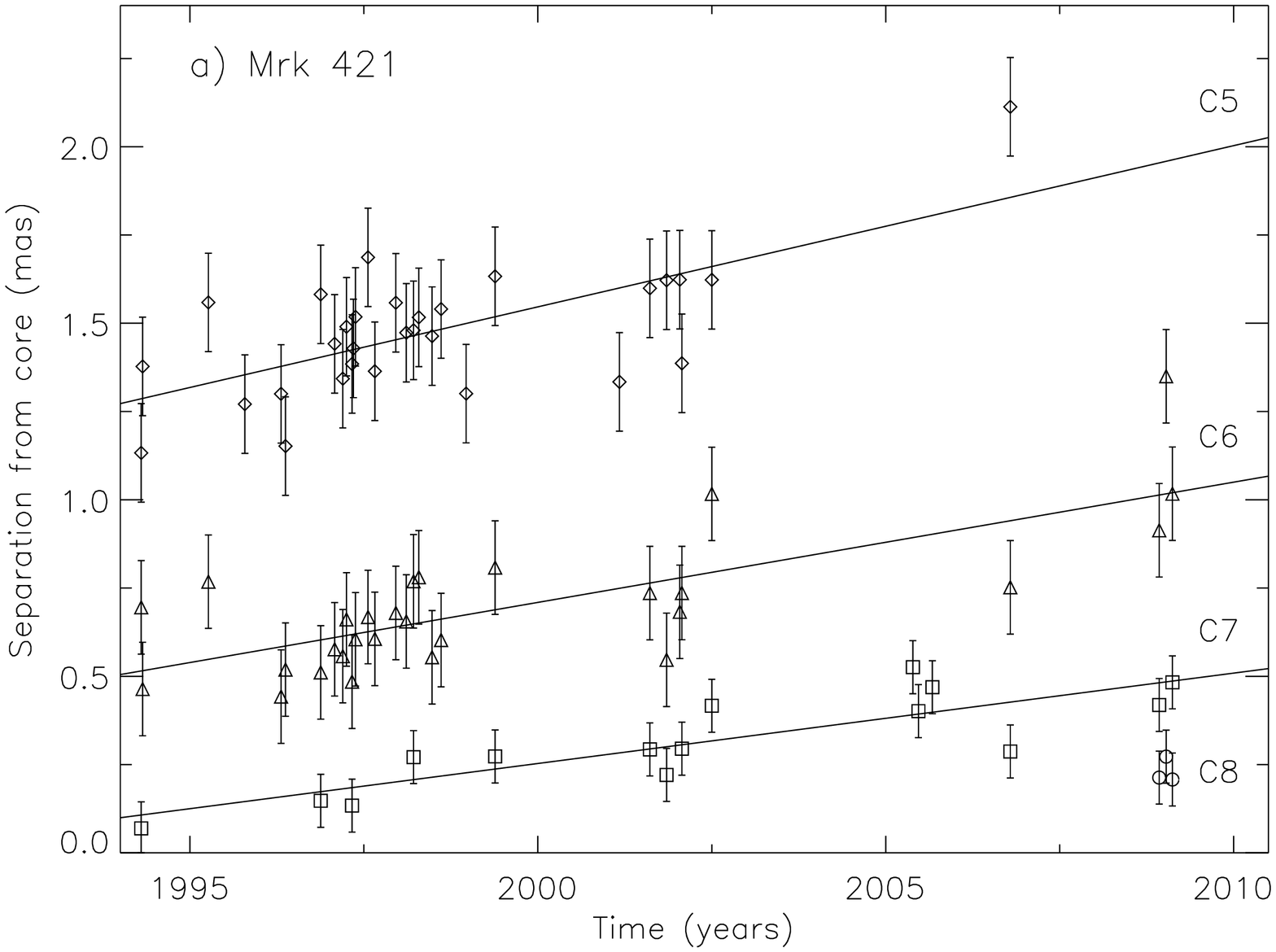}
\includegraphics[angle=90,scale=0.45]{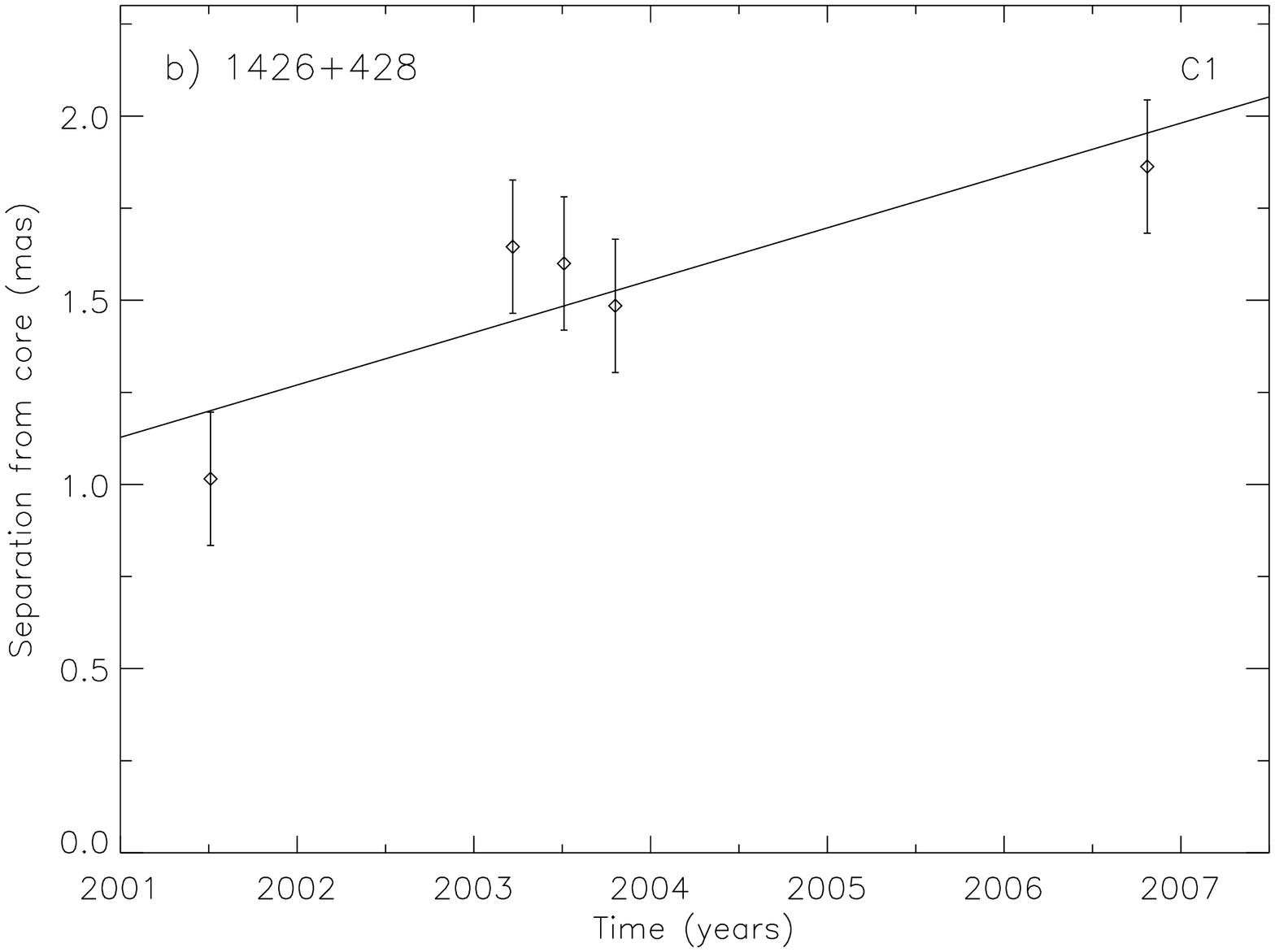}
\caption{Separations from the core of Gaussian model components 
versus time. Lines are the least-squares fits to outward motion with constant speed.
{\em (a)}: Mrk~421. {\em (b)}: H~1426+428. {\em (c)}: Mrk~501. 
{\em (d)}: 1ES~1959+650. {\em (e)}: PKS~2155$-$304. {\em (f)}: 1ES~2344+514.}
\end{center}
\end{figure*}

\begin{figure*}
\begin{center}
\includegraphics[angle=90,scale=0.45]{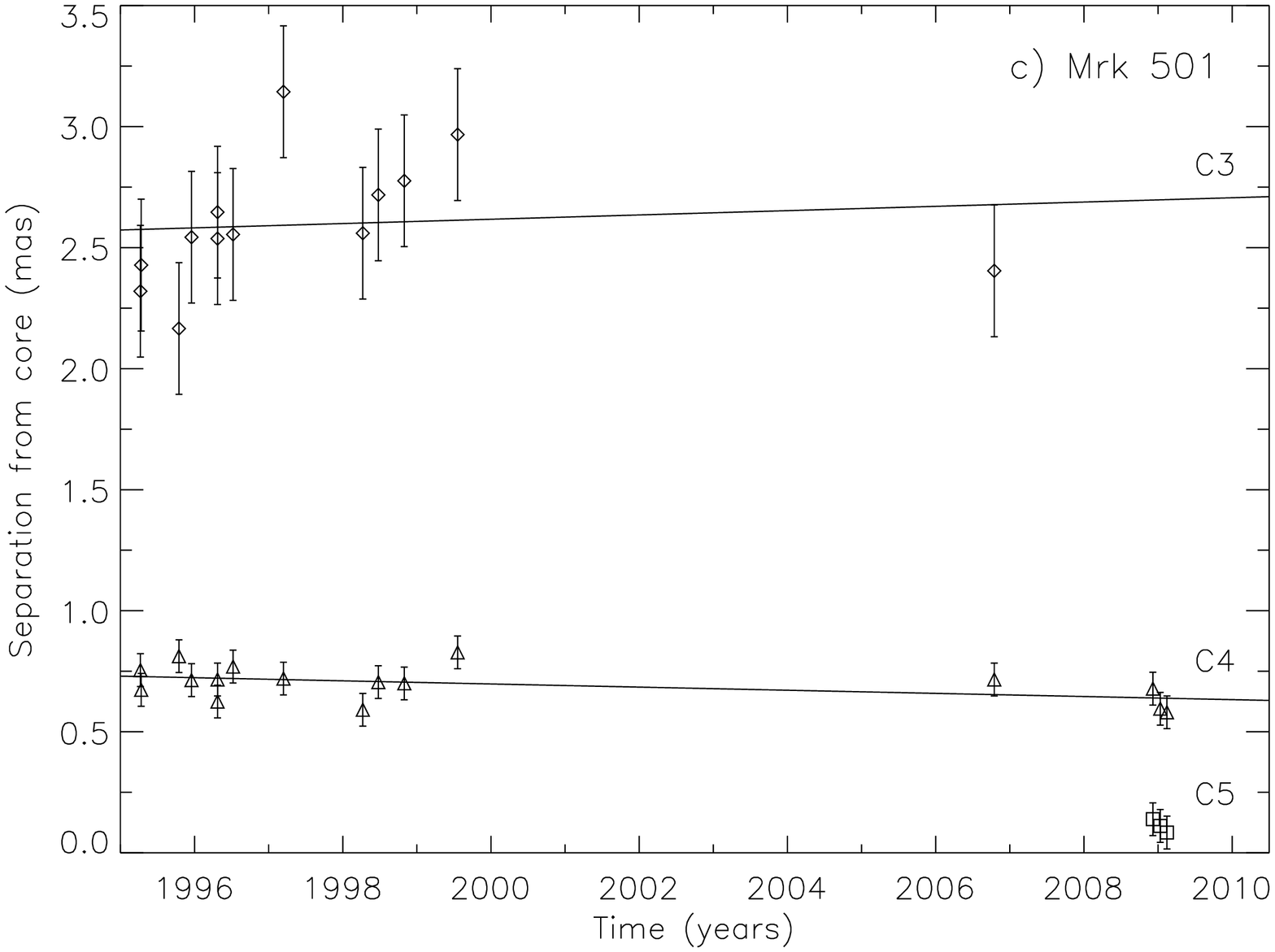}
\includegraphics[angle=90,scale=0.45]{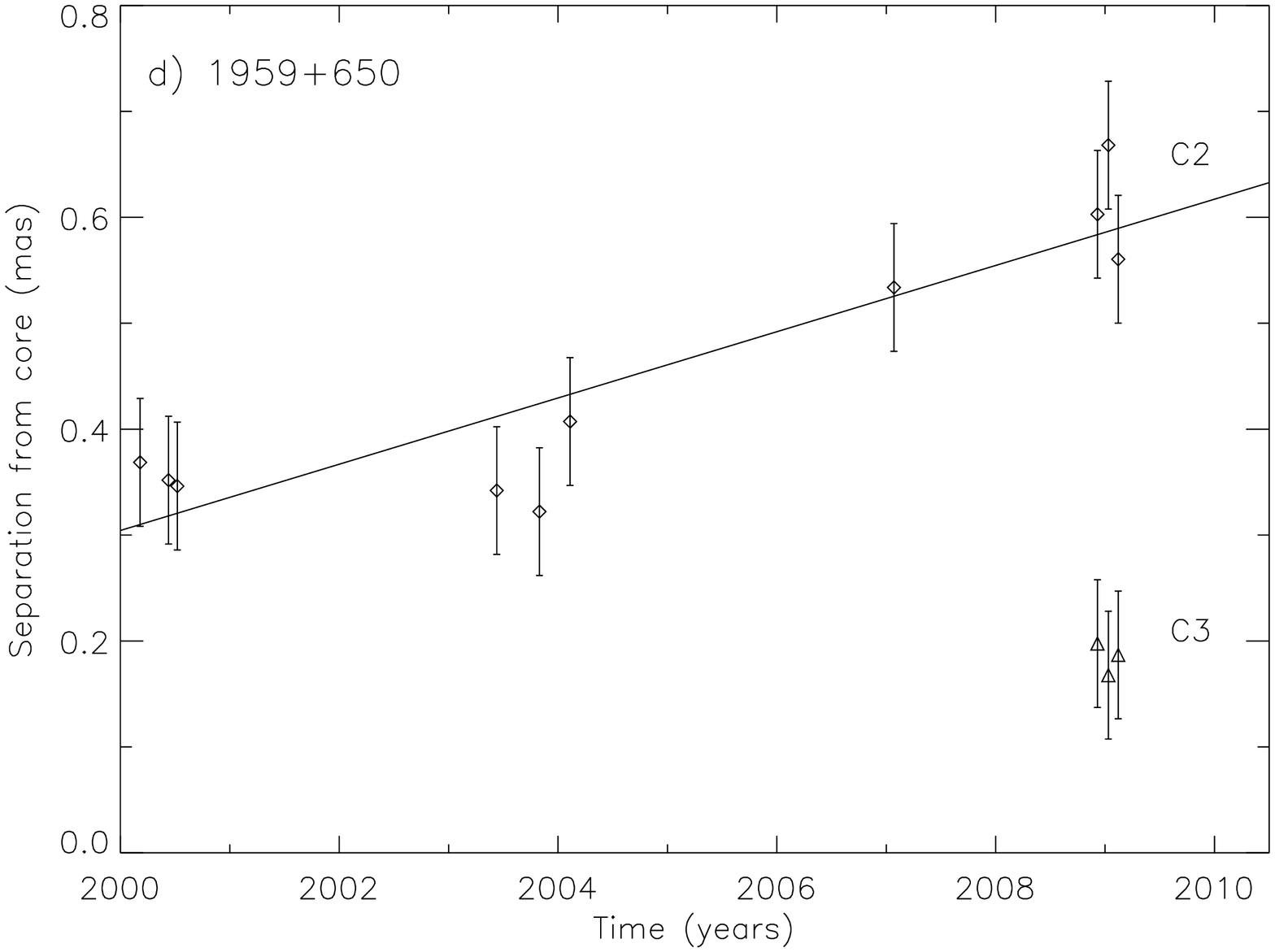}
\end{center}
\begin{center}
{\bf Figure. 13.}---{\em Continued}
\end{center}
\end{figure*}

\begin{figure*}
\begin{center}
\includegraphics[angle=90,scale=0.45]{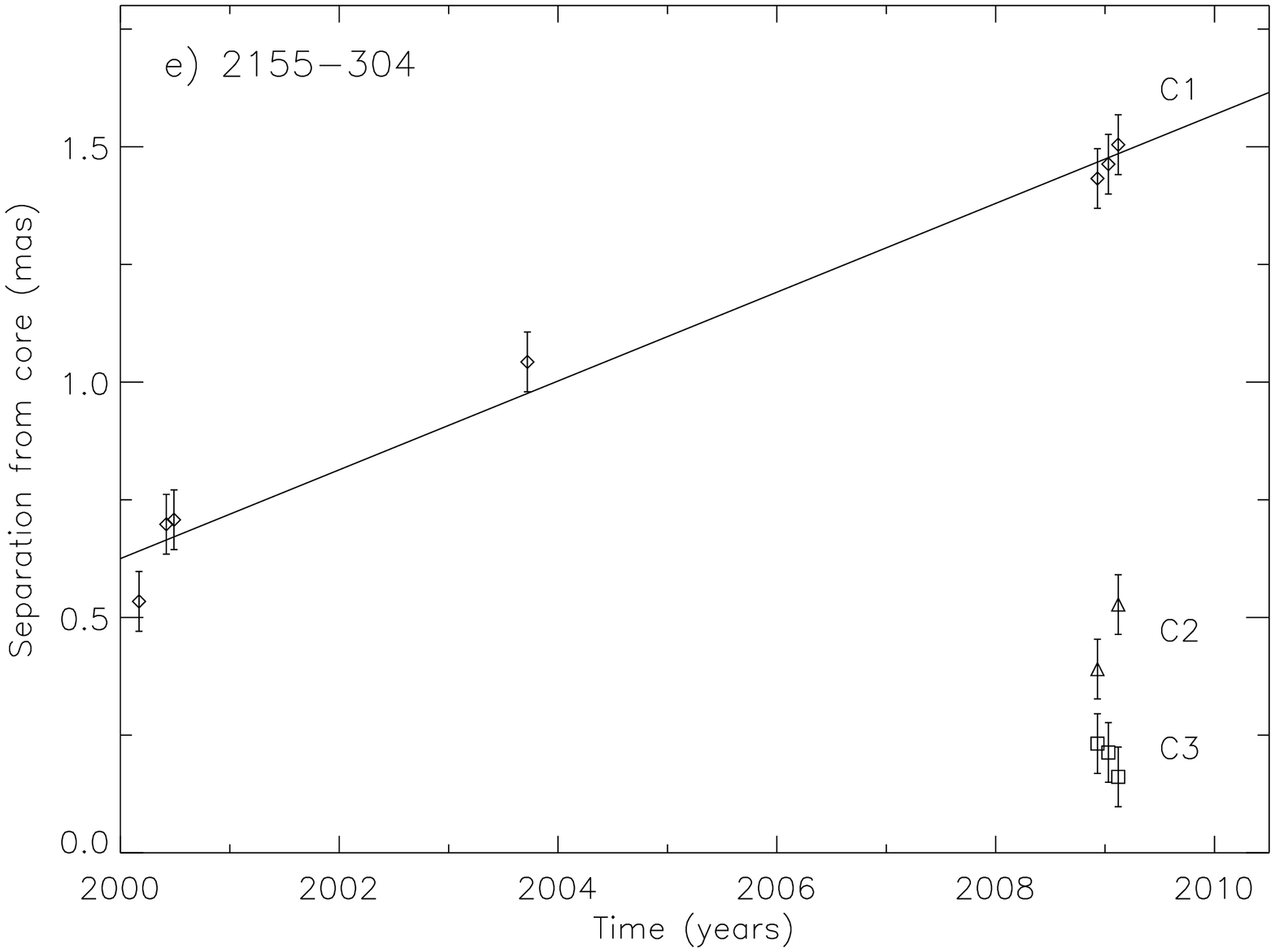}
\includegraphics[angle=90,scale=0.45]{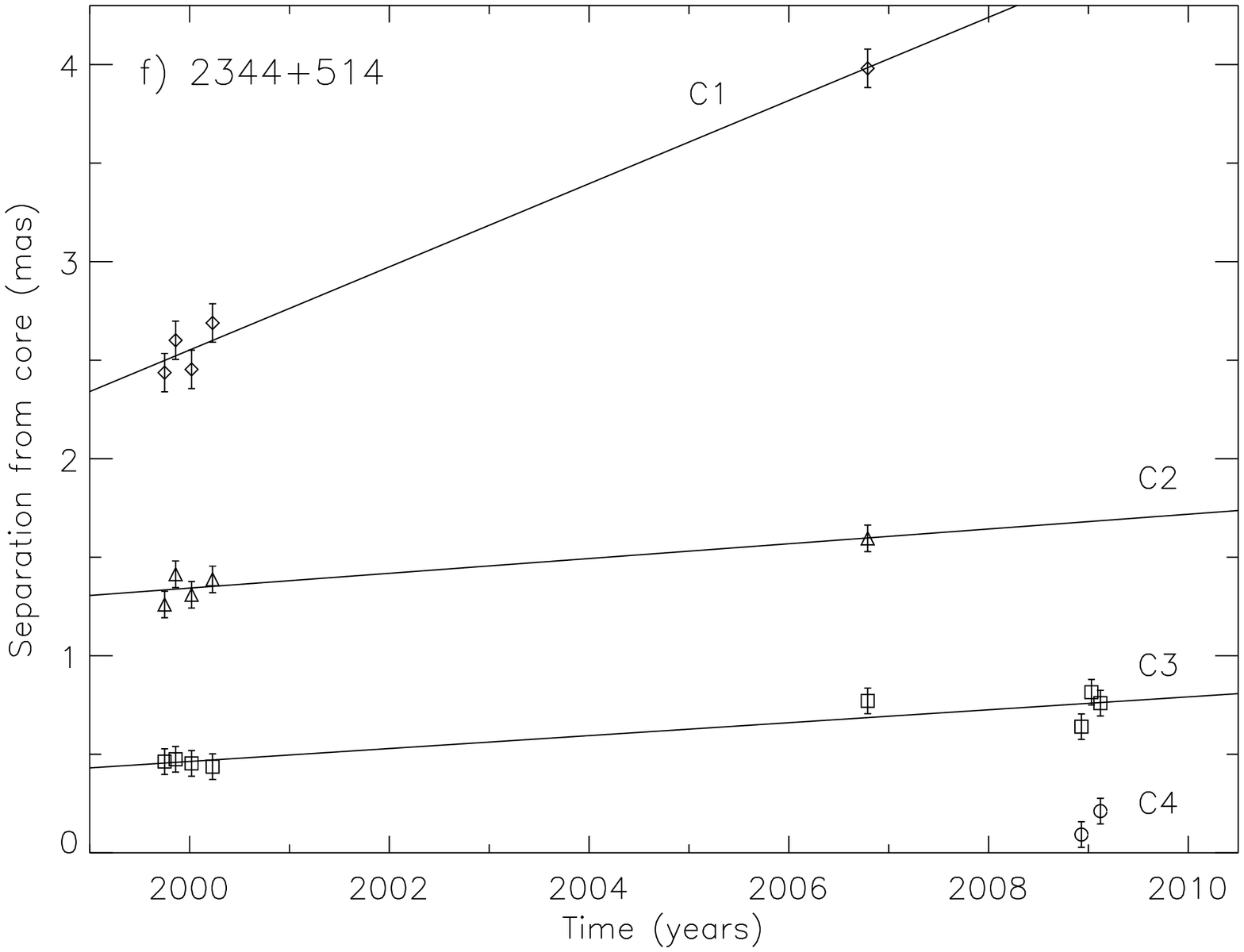}
\end{center}
\begin{center}
{\bf Figure. 13.}---{\em Continued}
\end{center}
\end{figure*}

Apparent speeds were measured from linear least-squares fits to the separation versus time plots.
The measured apparent speeds are tabulated in Table~\ref{speedtab}.
This table updates the similar table from P08 (Table~4~in that paper),
which in turn updated Table~3 from Piner \& Edwards (2004).
Table~\ref{speedtab} includes apparent speed measurements for a total of 22 Gaussian components.
Six of these components are new inner components detected only in this paper, their measured speeds
are typically upper limits and thus fits are not shown for them in Figure~13; these upper limits are discussed
further in the next subsection. Five of the components appear only in lower frequency images and thus are not
detected in this paper or shown on Figure~13, but they are included in Table~\ref{speedtab} for completeness.
The remaining eleven components are detected both in earlier work and this paper, and Table~\ref{speedtab}
provides updated and more precise measurements of their apparent speeds. The apparent speed fits for those
eleven components are shown in Figure~13. Although there are time gaps of several years between
observations of particular sources, these sources have only a few slowly moving jet components,
so that there are no ambiguities in identifying components between epochs.

\begin{table*}
\caption{Apparent Component Speeds in TeV HBLs}
\vspace{-0.15in}
\begin{center}
\label{speedtab}
{\small \begin{tabular}{l l r r c c} \colrule \colrule \\ [-9pt]
& & & & & Fastest Significant\\
& & & \multicolumn{1}{c}{Apparent Speed} & & Apparent Speed \\
\multicolumn{1}{c}{Source} & \multicolumn{1}{c}{Comp.} 
& \multicolumn{1}{c}{$N$} & \multicolumn{1}{c}{(multiples of $c$)}  
& Ref. & (multiples of $c$) \\ 
\multicolumn{1}{c}{(1)} & \multicolumn{1}{c}{(2)} & \multicolumn{1}{c}{(3)} &
\multicolumn{1}{c}{(4)} & (5) & (6) \\ \colrule \\ [-5pt]
Mrk~421         & C4  & 20 & $0.09\pm0.09$  & 1 & $0.09\pm0.02$ \\
                & C4a & 14 & $-0.09\pm0.09$ & 1 &               \\
                & C5  & 30 & $0.09\pm0.02$  & 4 &               \\
                & C6  & 29 & $0.07\pm0.01$  & 4 &               \\
                & C7  & 15 & $0.05\pm0.01$  & 4 &               \\
                & C8  & 3  & $-0.03\pm1.12$ & 4 &               \\
H~1426+428      & C1  & 5  & $1.20\pm0.40$  & 4 & $1.20\pm0.40$ \\
Mrk~501         & C1  & 11 & $0.07\pm0.15$  & 2 & $0.47\pm0.10$ \\
                & C2  & 9  & $0.47\pm0.10$  & 2 &               \\
                & C3  & 13 & $0.02\pm0.06$  & 4 &               \\
                & C4  & 16 & $-0.01\pm0.01$ & 4 &               \\
                & C5  & 3  & $-0.66\pm1.14$ & 4 &               \\
1ES~1959+650    & C1  & 3  & $-0.14\pm0.75$ & 3 & $0.10\pm0.02$ \\
                & C2  & 10 & $0.10\pm0.02$  & 4 &               \\
                & C3  & 3  & $-0.19\pm1.40$ & 4 &               \\
PKS~2155$-$304  & C1  & 7  & $0.72\pm0.05$  & 4 & $0.72\pm0.05$ \\
                & C2  & 2  & $5.48\pm3.59$  & 4 &               \\
                & C3  & 3  & $-2.80\pm3.58$ & 4 &               \\
1ES~2344+514    & C1  & 5  & $0.62\pm0.05$  & 4 & $0.62\pm0.05$ \\
                & C2  & 5  & $0.11\pm0.03$  & 4 &               \\
                & C3  & 8  & $0.10\pm0.02$  & 4 &               \\
                & C4  & 2  & $1.83\pm1.41$  & 4 &               \\ \colrule
\end{tabular}}
\end{center}
NOTES.--- Col.~(3): 
Number of observations of component.
Col.~(6): Fastest apparent speed with at least 2$\sigma$ significance.
References for VLBI data. --- (1) Piner \& Edwards (2005). (2) Edwards \& Piner (2002)
with modified cosmological parameters. (3) Piner \& Edwards (2004).
(4) This paper. 
\end{table*}

Figure~13 and Table~\ref{speedtab} combine measurements at different observing
frequencies, so can potentially be affected by frequency-dependent separation of components from the VLBI core,
caused by the varying opacity of the core with frequency.
We previously found this effect to be small in both Mrk~421 (Piner et al. 1999) and 
1ES~1959+650 (P08). Croke et al. (2010) find measurable frequency-dependent
offsets in Mrk~501 at lower frequencies, but their fit to this effect shows that it
should be negligible at higher (22 to 43~GHz) frequencies. 
Our later observing epochs are generally at higher frequencies, in which case frequency-dependent separation
would tend to move the later points out, increasing the measured speeds.
If frequency-dependent separation is significant, then the speeds in Table~\ref{speedtab}
are upper limits. 
We also note that while the representation of the jet with Gaussian model components 
does not reproduce all of the details of the
complex morphology of the jet images, it should reflect the centroid of any bulk
outward motion of the jet plasma.

The statistical results of the apparent speed measurements in Table~\ref{speedtab}
are consistent with those found in earlier papers, but now with more precise values for
the apparent speeds in individual sources. In particular, all eleven newly updated speeds of
previously identified components are statistically consistent
with their previously measured values from P08
within about $2\sigma$ for all components; the largest outlier is a $2.2\sigma$ difference
for the measured speed of component C2 in 1ES~1959+650 between Table~\ref{speedtab} in this paper and P08.
The sources that have the greatest reduction in the errors associated with their apparent speeds are
PKS~2155$-$304 and 1ES~2344+514, by factors exceeding six in both cases compared to the values
given in P08. This is due to the significantly
expanded time baseline for observations of these two sources and the improvement in resolution
by observing at 43~GHz.

The most important column in Table~\ref{speedtab} is the final column giving the fastest 
measured apparent speed in each blazar.
Although patterns in the flow do sometimes remain stationary or move slower than the bulk speed,
for the majority of sources, the peak apparent speed of model components
measured over many years of VLBI monitoring is indicative
of the bulk apparent speed (e.g., Cohen et al. 2007).
With the exception of the two very slow sources (Mrk~421 and 1ES~1959+650), the mean peak speed
from Table~\ref{speedtab} is $0.8c$, consistent with the $1c$ typical peak speed given in P08.
The low VLBI apparent speeds are but one part of a
convergence of multiple lines of evidence indicating modest bulk Lorentz factors in the parsec-scale radio jets
of HBLs (P08 and references therein).
These apparent speeds are related to the bulk Lorentz factors of the radio jets in the following subsection.

\subsubsection{Limits on Motions near the Core}
\label{limits}
Since some proposed solutions to the bulk Lorentz factor crisis 
for TeV blazars (see $\S$~\ref{intro}) invoke a decelerating jet, it is important
to measure the jet apparent speeds as close to the central engine as possible,
in order to either directly observe or place limits on the length scale of the putative deceleration.
The 43~GHz VLBA observations presented in this paper were made in part for this
particular goal. Fortunately,
components near the core at 43~GHz were bright and could be easily distinguished from the core
in the Gaussian model fitting,
and new model components were detected in the 43~GHz datasets 
for all five sources at about 0.1-0.2~mas from the core (see Table~\ref{mfittab}).

Because the data on these new components span only two months, most measured apparent speeds
in Table~\ref{speedtab} for these new components are only upper limits.
In Table~\ref{limittab}, we present the data for the innermost component in each source in
the form of a $2\sigma$ upper limit to the apparent speed (or a measured value, if available). 
Because measured black hole masses are now
available for many TeV blazars (e.g., Wagner [2008] and references therein), 
the separation of these innermost components from the core
can also be expressed in terms of Schwarzschild radii ($R_{S}$), which is done in Table~\ref{limittab}.
The measured VLBI separations are deprojected using an assumed viewing angle of $2.5\arcdeg$, which
is the median viewing angle listed for these six blazars in the analysis by
Celotti \& Ghisellini (2008). If jet bending away from the line-of-sight occurs 
(as has been postulated for Mrk~421 and Mrk~501, see $\S$~\ref{limb}),
then the deprojected numbers in terms of parsecs or Schwarzschild radii are upper limits.

\begin{table*}[!t]
\caption{Upper Limits on Apparent Speeds of Innermost Components}
\vspace{-0.15in}
\begin{center}
\label{limittab}
{\small \begin{tabular}{l l c r r r c c} \colrule \colrule \\ [-9pt]
& & Upper Limit & \multicolumn{1}{c}{$r$} & \multicolumn{1}{c}{$r$} &
\multicolumn{1}{c}{$r_{d}$} & $M_{BH}$ & $r_{d}$ \\
\multicolumn{1}{c}{Source} & \multicolumn{1}{c}{Comp.} & ($c$) &
\multicolumn{1}{c}{(mas)} & \multicolumn{1}{c}{(pc)} & \multicolumn{1}{c}{(pc)} & 
($10^{9}M_{\odot}$) & ($10^{5}R_{S}$) \\
\multicolumn{1}{c}{(1)} & \multicolumn{1}{c}{(2)} & (3) &
\multicolumn{1}{c}{(4)} & \multicolumn{1}{c}{(5)} & \multicolumn{1}{c}{(6)} &
(7) & (8) \\ 
\colrule \\ [-5pt]
Mrk~421        & C8 & 2.2 & 0.23 & 0.14 & 3.1  & 0.36 & 0.9  \\
H~1426+428     & C1 & 1.2 & 1.52 & 3.46 & 79.4 & 0.45 & 18.4 \\
Mrk~501        & C5 & 1.6 & 0.11 & 0.07 & 1.7  & 1.00 & 0.2  \\
1ES~1959+650   & C3 & 2.6 & 0.18 & 0.16 & 3.8  & 0.15 & 2.6  \\
PKS~2155$-$304 & C3 & 4.4 & 0.20 & 0.42 & 9.5  & 0.96 & 1.0  \\
1ES~2344+514   & C4 & 1.8 & 0.15 & 0.13 & 2.9  & 1.10 & 0.3  \\ \colrule
\end{tabular}}
\end{center}
NOTES.--- Col.~(3): Two sigma upper limit on innermost component speed, or measured
value if not an upper limit. Col.~(4): Average separation of innermost component
from the core, in mas. Col.~(5): Average separation of innermost component, in parsecs.
Col.~(6): Deprojected separation of innermost component, assuming a
viewing angle of $2.5\arcdeg$. Col.~(7): Black hole mass from Wagner (2008).
Col.~(8): Deprojected separation of innermost component in Schwarzschild radii.
\end{table*}

The results in Table~\ref{limittab} show that the median upper limit to the apparent jet
speed is 2$c$ at a median distance of $10^{5}R_{S}$. 
We can use these results to place a limit on the bulk Lorentz factor in the 43~GHz VLBI
jet ($\Gamma_{2}$) relative to the bulk Lorentz factor 
in the TeV emitting region ($\Gamma_{1}$), assuming that the limits
on the VLBI proper motions place a limit on the bulk motion of the plasma,
and following a line of argument originally from Stern \& Poutanen (2008).
Using the standard formula for apparent superluminal speed, $\beta_{app}=\beta\sin\theta/(1-\beta\cos\theta)$,
and the binomial expansion, then $\beta_{app}\approx 2\Gamma_{2}^{2}\theta$, where we assume that
since we are observing a TeV-selected sample, we have a viewing angle of $\theta\approx 1/\Gamma_{1}<<1/\Gamma_{2}$. 
Inserting this viewing angle, and the observed limit $\beta_{app}<2$, we obtain
$2\Gamma_{2}^{2}/\Gamma_{1}<2$, or $\Gamma_{2}<(\Gamma_{1})^{1/2}$, at $10^{5}R_{S}$.
In this case, the $\Gamma_{1}$ and $\Gamma_{2}$ refer to the TeV-emitting and 43~GHz-emitting
regions irrespective of the specific geometry, as long as their velocity vectors are aligned.
So the two Lorentz factors may refer to
an inner and an outer jet (e.g., Georganopoulos \& Kazanas 2003), 
a fast spine and a slow sheath (e.g.,  Ghisellini et al. 2005), 
a fast `needle' embedded in a slower jet (Ghisellini \& Tavecchio 2008),
or the leading and trailing regions of a non-stationary ejection (Lyutikov \& Lister 2010).
\footnote{
This analysis would not apply to the `jet in a jet' model by Giannios et al. (2009), because
in that model the TeV emitting region can move in a different direction from the main jet.}
We note that farther out in the VLBI jets where peak apparent speeds of about $1c$ are measured,
the corresponding limit is $\Gamma_{2}<(\Gamma_{1}/2)^{1/2}$, at about $10^{6}R_{S}$. 
For a typical Lorentz factor derived from the TeV SED modeling of $\Gamma_{1}\sim25$, this implies
$\Gamma_{2}<\sim5$ and $\Gamma_{2}<\sim3.5$ at $10^{5}$ and $10^{6}R_{S}$, respectively.

Such limits are just consistent with lower limits on the Lorentz factor 
in TeV blazars derived from measured lower limits on the VLBI jet to counterjet brightness ratio.
The most stringent of these limits comes from High-Sensitivity Array (HSA)
observations of Mrk 501, which constrain $\beta\cos\theta>0.92$ in that source,
corresponding to $\Gamma>\sim3$ for viewing angles of a few degrees (Giovannini et al. 2008).
These limits are also consistent with some specific jet models for TeV blazars. 
For example, the decelerating jet model by Levinson (2007)
posits radiative deceleration of emitting blobs to Lorentz factors of a few
on scales of less than about $10^{3}$ Schwarzschild radii.
In the non-stationary ejection model by Lyutikov \& Lister (2010),
the ratio of the Lorentz factor of the leading edge to that of the bulk flow can
be as high as $2\sigma^{2/3}$, where $\sigma$ is the plasma magnetization, 
and this can reach tens for highly magnetized flows (the limits given above imply 
$\sigma>\sim4$ to 7~in the context of their model).
The limits given above are also consistent with models where 
the VLBI jet emission is dominated by a slower sheath surrounding
a faster spine; this hypothesis also matches well with the transverse structures 
in both total intensity and linear polarization described in the previous section,
provided that such a spine-sheath structure is already established by distances of
$10^{5}R_{S}$ from the central engine.
Given that such transverse structures are here directly observed in 
the jets of three of the closer TeV blazars,
we argue that these transverse structures probably play an important role in TeV blazar jet physics, and may
themselves influence, or be influenced by, a radially decelerating jet, as in the models by Ghisellini et al. (2005)
and Stern \& Poutanen (2008).

\section{Conclusions}
\label{conclusion}
We have presented a total of 23 new VLBA images, most at 43~GHz, of 
the first six blazars to be well established as TeV sources.
Below we enumerate some specific observational conclusions from the present work, and remark
on their significance.
\begin{enumerate}
\item{This paper has presented the first published 43~GHz images for three of these sources
(1ES~1959+650, PKS~2155$-$304, and 1ES~2344+514), and they show these jets at 
factors of a few higher resolution than has previously been attained.
These high-resolution images reveal some new morphological features, such as strong
bending in the jet of PKS~2155$-$304.
This establishes strong jet bending on VLBI scales as a common property of TeV blazars,
implying viewing angles close to the line-of-sight.}
\item{Significant limb-brightening is detected in inner milliarcsecond of Mrk~421,
and analysis shows that it is very similar to the limb-brightening seen previously in Mrk~501
($\S$~\ref{limb}).}
\item{Transverse polarization structures are present in
the EVPA and fractional polarization images of Mrk~421, Mrk~501, and 1ES~1959+650.
The structures take the form of
a parallel EVPA along the jet axis, and an orthogonal EVPA at
the jet edges, along with an increase in fractional polarization at the jet edges.
Such transverse polarization structures may directly relate to the intrinsic magnetic field
geometry as, e.g., a large-scale helical magnetic field in a resolved cylindrical
jet ($\S$~\ref{transversepol}).}
\item{New measured positions of previously known jet components are used
to update apparent jet speed measurements; in many cases the expanded time baseline
has significantly reduced the statistical error compared to previously published results
($\S$~\ref{update}).}
\item{We detect new jet components in the 43~GHz images of these sources at distances of
0.1-0.2 mas from the core.
We place upper limits on the apparent speeds of these components of
$<2c$. From these limits, we conclude that
$\Gamma_{2}<(\Gamma_{1})^{1/2}$, at $\sim10^{5}$ Schwarzschild radii, where
$\Gamma_{1}$ and $\Gamma_{2}$ are the bulk Lorentz factors in the TeV-emitting and 43~GHz-emitting
regions, respectively, assuming that their velocity vectors are aligned.
This implies bulk Lorentz factors in the 43~GHz-emitting regions of $\Gamma_{2}<\sim5$,
for typical Lorentz factors derived from the TeV SED modeling of $\Gamma_{1}\sim25$
($\S$~\ref{limits}).}
\end{enumerate}

It is significant that the three of these six TeV blazars that are both close enough and bright enough
to have their transverse structures resolved by the VLBA at 43~GHz (Mrk~421, Mrk~501, and 1ES~1959+650) all
show similar transverse structures in their total intensity, EVPA, and fractional polarization distributions.
It is also significant that new components detected close to the core in all of these sources have
apparent speeds of $<2c$, implying similar limits on the bulk Lorentz factor at similar distances
from the core in all six sources.
Taken together, these new observational results using the full resolution of the VLBA 
strongly favor theoretical models of TeV blazars that include significant transverse jet structures,
and where a modest Lorentz factor for the radio-emitting plasma is attained by $\sim10^{5}$ Schwarzschild radii
from the central engine.

\acknowledgments
We thank the anonymous referee for helpful comments that improved the paper.
The National Radio Astronomy Observatory is a facility of the National
Science Foundation operated under cooperative agreement by Associated Universities, Inc.
This research has made use of the NASA/IPAC Extragalactic Database (NED) 
which is operated by the Jet Propulsion Laboratory, California Institute of Technology, 
under contract with the National Aeronautics and Space Administration.
This work was supported by the National Science Foundation under Grant 0707523.

{\it Facilities:} \facility{VLBA ()}

\end{document}